\documentclass[superscriptaddress,aps,onecolumn,12pt,floatfix,altaffilletter,showpacs,showkeys,notitlepage,nofootinbib,preprintnumbers]{revtex4-2}

\usepackage[dvipdf,dvips,dvipdfmx]{graphicx}
\usepackage[colorlinks=true,citecolor=blue,linkcolor=blue]{hyperref}
\usepackage{amsmath}
\usepackage{bm}
\usepackage{amssymb}
\usepackage{epsfig}
\usepackage{epstopdf}
\usepackage{url}
\usepackage{color}
\usepackage[utf8]{inputenc} 

\usepackage{amsmath}
\usepackage{amssymb}
\usepackage{slashed}
\usepackage{cancel}
\usepackage{textcomp}
\usepackage{calc}
\usepackage{color}
\usepackage[utf8]{inputenc}

\allowdisplaybreaks

\catcode`\@=11
\def\lsim{\mathrel{\mathpalette\@versim<}}
\def\gsim{\mathrel{\mathpalette\@versim>}}
\def\@versim#1#2{\vcenter{\offinterlineskip
\ialign{$\m@th#1\hfil##\hfil$\crcr#2\crcr\sim\crcr } }}
\catcode`\@=12

\begin{document}

\newcommand{\bi}{\bibitem}
\newcommand{\change}[1]{{\color{red}#1}}
\newcommand{\vect}[1]{\bm{#1}}
\newcommand{\Slash}[1]{{\ooalign{\hfil/\hfil\crcr$#1$}}} 
\newcommand{\dcsb}{D$\chi$SB}
\newcommand{\Nf}{N_{\rm f}}
\newcommand{\Nc}{N_{\rm c}}
\newcommand{\qL}{q_\text{L}}
\newcommand{\qLone}{q_{\text L1}}
\newcommand{\qLtwo}{q_{\text L2}}
\newcommand{\tL}{t_\text{L}}
\newcommand{\bL}{b_\text{L}}
\newcommand{\tR}{t_\text{R}}
\newcommand{\e}{{\rm e}}
\renewcommand{\L}{\text{L}}
\newcommand{\R}{\text{R}}
\newcommand{\A}{\mathcal{A}}
\newcommand{\D}{\mathcal{D}}
\newcommand{\F}{\mathcal{F}}
\newcommand{\mP}{\mathcal{P}}
\newcommand{\ul}{\ensuremath{\underline}}
\newcommand{\ub}{\underbrace}
\newcommand{\ob}{\overbrace}
\newcommand{\zb}{\bar z}
\newcommand{\al}[1]{\begin{align}#1\end{align}}
\newcommand{\bp}{\begin{pmatrix}}
\newcommand{\ep}{\end{pmatrix}}
\newcommand{\nn}{\nonumber}
\newcommand{\paren}[1]{\left(#1\right)}
\newcommand{\sqbr}[1]{\left[#1\right]}
\newcommand{\br}[1]{\left\{#1\right\}}
\newcommand{\zero}{}
\newcommand{\one}{\I_2}
\newcommand{\gamhat}{\hat\gamma}
\newcommand{\ig}{ig}
\newcommand{\I}{\text{I}}
\newcommand{\gc}{g_{\rm c}}
\newcommand{\tb}{\textbf}
\newcommand{\p}{\partial}
\newcommand{\edh}{\text{\dh}}
\newcommand{\ho}{\hat{o}}
\newcommand{\red}{\textcolor{red}}
\newcommand{\te}{\text}
\newcommand{\ns}{{N\atop S}}
\newcommand{\sn}{{S\atop N}}
\newcommand{\N}{\hat{N}}
\newcommand{\GeV}{\,\text{GeV}}
\newcommand{\TeV}{\,\text{TeV}}
\newcommand{\cred}[1]{{\color{red}#1}}
\newcommand{\cblue}[1]{{\color{blue}#1}}
\newcommand{\df}{\text{d}}
\newcommand{\ab}[1]{\left|#1\right|}
\newcommand{\bs}[1]{\boldsymbol}
\newcommand{\X}{\mathcal X}
\newcommand{\pmat}[1]{\begin{pmatrix}#1\end{pmatrix}}
\newcommand{\bmat}[1]{\begin{bmatrix}#1\end{bmatrix}}
\newcommand{\fn}[1]{\!\left(#1\right)}
\newcommand{\bra}{\langle}
\newcommand{\ket}{\rangle}
\newcommand{\Lag}{\mathcal L}
\newcommand{\tr}{\text{tr}}
\newcommand{\im}{{\rm i}}
\newcommand\normalorder[1]{{:}\mkern1mu#1\mkern1.6mu{:}}
\newcommand{\be}{\begin{eqnarray}}
\newcommand{\ee}{\end{eqnarray}}

\newcommand{\Y}[1]{\textcolor{red}{(Y:#1)}} 




\title{Inflation and dark matter after spontaneous Planck scale generation by hidden chiral symmetry breaking}


\author{Mayumi \surname{Aoki}}
\email{mayumi.aoki@staff.kanazawa-u.ac.jp}
\affiliation{Institute  for  Theoretical  Physics,  Kanazawa  University,  Kanazawa  920-1192,  Japan}

\author{Jisuke \surname{Kubo}}
\email{kubo@mpi-hd.mpg.de}
\affiliation{Max-Planck-Institut f\"ur Kernphysik (MPIK), Saupfercheckweg 1, 69117 Heidelberg, Germany}
\affiliation{Department of Physics, University of Toyama, 3190 Gofuku, Toyama 930-8555, Japan}

\author{Jinbo \surname{Yang}}
\email{j$_$yang@hep.s.kanazawa-u.ac.jp}
\affiliation{Institute  for  Theoretical  Physics,  Kanazawa  University,  Kanazawa  920-1192,  Japan}
\pagestyle{plain}
\preprint{KANAZAWA-21-10}

%

\begin{abstract}
Dynamical chiral symmetry breaking in a QCD-like hidden sector
is used to generate
 the Planck mass and the electroweak scale
 including the heavy right-handed neutrino mass.
 A real scalar field transmits
 the energy scale of the hidden sector to the visible sectors,
playing besides a role of inflaton in the early Universe
while 
realizing  a Higgs-inflation-like model.
 Our dark matter candidates are hidden pions that raise
 due to dynamical chiral symmetry breaking. 
 They are produced from the decay of inflaton.
 Unfortunately, it will be impossible to directly 
detect them, because they  are super heavy ($10^{9\,\sim\,12}$ GeV),
and moreover the interaction with the visible sector is 
extremely suppressed.

\end{abstract}
%
%
\maketitle





%

\section{Introduction}
While the standard model (SM) has been successful in explaining a large number of experimental data,
the model fails to answer several questions, such as the neutrino mass, dark matter (DM),
matter-antimatter asymmetry, and inflation.
Neutrino masses cannot be explained with only the field content of the SM;
there are no viable candidates for DM; asymmetry of matter and antimatter in our Universe has been addressed for a long time as a serious problem; 
cosmic inflation provides a compelling explanation for the homogeneity and isotropy of the Universe
and for the observed spectrum of density perturbations.
However, the inflaton's nature remains unknown. 

Furthermore, the SM suffers from a hierarchy problem.
The only dimensionful parameter in the SM, the Higgs mass parameter $\mu_H$ is sensitive 
to quantum corrections, if there exist  high intermediate scales.
The correction $\delta \mu_H^2$ becomes huge as $\delta \mu_H^2\sim \Lambda^2$, where $\Lambda$ represents 
the high intermediate scales.
This implies that a sizable fine-tuning between various contributions to the Higgs mass is needed. 

Similarly to the SM case, in which $\mu_H$ is the only dimensionful parameter,
Einstein’s theory of gravity contains 
a single dimensionful parameter, the  Planck  mass (apart from the cosmological constant), and classical scale invariance forbids the presence
of these  dimensionful parameters in the Lagrangian.
Thus our  guiding principle to construct a model in this paper is scale invariance.
	Though scale invariance is anomalous, anomaly can not directly generate
	mass gaps. Consequently, scale invariance has to be spontaneously 
broken to generate a  mass gap - a physical energy scale.
	One possible method to generate a scale is 
the dynamical way by the Coleman-Weinberg potential~\cite{Coleman:1973jx}. 
Another possibility is the non-perturbative way
in nonabelian gauge theories, e.g., Quantum Chromodynamics (QCD)~\cite{Nambu:1960xd,Nambu:1961tp,Nambu:1961fr}. 
In the present paper we  ask ourselves whether it is possible
	to generate the Planck mass and  the electroweak scale from
	the same origin. 
	In this way  a new view on the hierarchy problem
	may be gained, because the hierarchy problem is a problem
	among different mass scales.
%
%
%
%
 
 If one is to maintain classical scale invariance
in a complete theory including the gravity, therefore,
it is necessary to generate the Planck  mass,
through the spontaneous breaking of the scale invariance~\cite{Brans:1961sx,Terazawa:1976eq, Akama:1977hr,Terazawa:1981ga}. 
This is realized in, e.g., the Brans-Dicke theory~\cite{Brans:1961sx} 
where the $M_\text{Pl}$ is generated dynamically by the vacuum expectation value (VEV) of a scalar field.
The non-minimal coupling term $S^2 R$ generates the effective  Einstein-Hilbert term,
where $S$ is a real scalar field and $R$ is the Ricci scalar.
This scenario incorporates inflation while  identifying the scalar 
field as inflaton
~\cite{Bezrukov_2008,Rinaldi_2016,Tambalo_2017,Ferreira_2016,Ferreira_2017,Ferreira_2019,Kubo:2018kho,Kubo:2020fdd,
Guth:1980zm,Linde:1981mu,Linde:1982zj,Albrecht:1982wi,Linde:2007fr,
Salvio:2014soa,Kannike:2015apa,Karam:2018mft,Ghilencea:2019rqj,Ghilencea:2018thl,Farzinnia:2015fka,Gialamas:2021enw}. 
In this case, the Planck mass is dynamically generated
during or before inflation.

In this paper, we construct a classical scale invariant model with $R+R^2$ gravity ~\cite{Starobinsky:1980te,1981JETPL..33..532M,Starobinsky:1983zz}
where the Planck mass is generated dynamically by the VEV of 
an inflaton.
The original $R^2$ inflation is known as the Starobinsky inflation~\cite{Starobinsky:1980te}, in which
 inflaton is the degree of freedom related to the $R^2$ term~\cite{Barrow:1988xh,Maeda:1988ab}.
Predictions of  the $R^2$ inflation, such as a spectral index $n_s$ and a tensor-to-scalar ratio $r$, fit observations 
in the Planck experiment~\cite{Aghanim:2018eyx,Akrami:2018odb} very well.
In  scale invariant extensions of the $R^2$ inflation model, on the other hand,
a double-field inflation system is realized due to the non-minimal coupling $S^2 R$~\cite{Guth:1980zm,Linde:1981mu,Linde:1982zj,Albrecht:1982wi,Linde:2007fr}.

We introduce a strongly interacting QCD-like hidden sector, in which the mass scale is generated in a non-perturbative way via condensation that breaks chiral symmetry dynamically.
The Higgs mass is generated by the quantum correction.
We apply the ``Neutrino option''~\cite{Brivio:2017dfq}, 
 where 
 $\mu_H^2$ is generated by the quantum correction of the right-handed neutrinos.
To obtain a consistent scale 
in the neutrino option scenario, 
the masses of the right-handed neutrinos are $\sim 10^{7}$ GeV~\cite{Vissani:1997ys,Casas:1999cd,Clarke:2015gwa,Bambhaniya:2016rbb}.
The Majorana mass term of the right-handed neutrino is dynamically generated via the chiral symmetry breaking~\cite{Meissner:2006zh,Brdar:2018vjq}, while 
the light active neutrinos obtain the masses through the type-I seesaw mechanism~\cite{Minkowski,Yanagida:1979as,GellMann:1980vs,Goran}.
In this model, there exist 
(quasi) Nambu-Goldstone (NG) bosons due to the dynamical chiral symmetry breaking,
and due to their stability they are good DM candidates
\cite{Hur:2011sv,Heikinheimo:2013fta,Holthausen:2013ota,Hatanaka:2016rek,Kubo:2014ida,Ametani:2015jla}.

The paper is organized as follows. In Section \ref{sec2},
we introduce the scale invariant model with a QCD-like hidden sector. 
We  use Nambu-Jona-Lasinio (NJL) model as an effective low-energy theory in a mean field approximation.
The masses of Planck, neutrinos, and Higgs boson are discussed. 
In Section \ref{sec3}, we show an effective action and discuss the inflationary dynamics in our model.
The results of the numerical study for the predicted inflation parameters are presented.
In Section \ref{sec4}, we discuss the DM candidates which are produced during or after the reheating phase.
The reheating temperature and the DM relic abundance are computed.
We summarize and conclude in Section \ref{sec5}.
\section{The Model} \label{sec2}
\label{Model}
Our model should describe  (i) the generation 
of a robust energy scale 
through the formation of chiral condensate,
(ii) gravity, (iii)
the SM interactions, (iv) including heavy right-handed neutrinos.
We impose scale invariance, and therefore no part of the model 
should contain any dimensionful parameter at the classical level.
\\
(i) In this part of the model, a robust energy scale is generated,
which is the unified origin of energy scales of  other sectors.
Instead of the Coleman-Weinberg mechanism \cite{Coleman:1973jx} 
which has been used  for the unified origin of energy scales in \cite{Kubo:2020fdd},
we employ here a strongly-interacting, QCD-like theory in 
which chiral symmetry is dynamically broken \cite{Nambu:1960xd,Nambu:1961tp,Nambu:1961fr}
and as a result a robust energy scale is generated. This part is described by
 \setcounter{equation}{0}
\renewcommand\theequation{2.\arabic{equation}}
\al{\frac{{\cal L}_H  }{\sqrt{-g}}&=	-\frac{1}{2}\mbox{Tr}~F^2+
\mbox{Tr}~\bar{\psi}(i \slashed{D} 
 -\bm{y} S)\psi \,,
\label{LH}}
where $F$ is the matrix-valued field strength tensor of the $SU(N_c)_H$ gauge theory, 
coupled with the vector-like hidden fermions $\psi_i~(i=1,\dots,n_f)$ belonging to the fundamental representation of  $SU(n_c)_H$,
and $S$ is a real SM singlet scalar. The strong dynamics of
the QCD-like theory (\ref{LH}) forms  a gauge invariant chiral condensate 
$\langle \bar{\psi} \psi\rangle$, which  generates
a linear term in $S$ in the potential, leading to a nonzero VEV
of $S$ \cite{Hur:2011sv, Heikinheimo:2013fta, Holthausen:2013ota, Hatanaka:2016rek}. The scalar $S$  plays a role of 
the mediator, because it subsequently  transfers the robust 
energy scale  to the other sectors of the model.
${\bm y}$ is an $n_f\times n_f$ Yukawa coupling matrix which can be assumed as a diagonal matrix without loss of generality, i.e. 
${\bm y}=\mbox{diag.}(y_1,\dots,y_{n_f})$.
This Yukawa coupling violates explicitly chiral symmetry.
Consequently, the NG bosons,
 associated with the dynamical chiral symmetry breaking
 $SU(n_f)_L\times SU(n_f)_R \to SU(n_f)_V$, acquire their masses and can become DM candidates due to the remnant unbroken flavor group $SU(n_f)_V$ (or its subgroup, depending on the choice of $y_i$) that can stabilise them \cite{Hur:2011sv, Heikinheimo:2013fta, Holthausen:2013ota, Hatanaka:2016rek}.
Note that due to the presence of the  fermions the use of the vierbein formalism is quietly understood. 
But it does not play any role in the following discussions.\\
(ii) This part is described by the Lagrangian 
\al{
\frac{{\cal L}_\text{G} }{\sqrt{-g}}  &=
-\frac{ \beta}{2}\,S^2\,R +\gamma\,R^2+
\kappa\,W_{\mu\nu\alpha\beta}W^{\mu\nu\alpha\beta} \,,
\label{LCGR}
}
where $R$ denotes the Ricci curvature scalar, and
$W_{\mu\nu\alpha\beta}$ is the Weyl tensor.
The Ricci curvature tensor squared,
$R_{\mu\nu\alpha\beta}R^{\mu\nu\alpha\beta}$,
is omitted in the Lagrangian (\ref{LCGR}), because 
it (and also $R_{\mu\nu}R^{\mu\nu}$)
can be written 
as a linear combination of $R^2$, $W_{\mu\nu\alpha\beta}W^{\mu\nu\alpha\beta} $ and 
the Gau\ss-Bonnet term which is  a surface term.
The non-minimal coupling 
$S^2\,R$ produces the Einstein-Hilbert term
when the real scalar field $S$ acquires the VEV denoted by  $v_S$,
with the (reduced) Planck mass $M_\text{Pl}\simeq \sqrt{\beta} v_S$.
Then the scalar $S$ and the scalaron $\varphi$ \cite{Starobinsky:1980te}, which describes the scalar degree 
of freedom related to the $R^2$ term, form a double-filed inflaton system to
describe cosmic inflation \cite{Guth:1980zm,Linde:1981mu,Linde:1982zj,Albrecht:1982wi,Linde:2007fr}.\\
(iii) The Lagrangian of this part consists of  the SM Lagrangian
with the Higgs mass term suppressed,   
$\left.{\cal L}_{\rm SM}\right|_{\mu_H=0}$, and also
of the Lagrangian for $S$: 
\al{\frac{{\cal L}_{{\rm SM}+S} }{\sqrt{-g}} &=\frac{ \left.{\cal L}_{\rm SM}\right|_{\mu_H=0} }{\sqrt{-g}}+\frac{1}{2}g^{\mu\nu}\partial_\mu S \partial_\nu S
-\frac{1}{4}\lambda_{HS} S^2 H^\dag H-\frac{1}{4} \lambda_S S^4\,,
\label{LSM}	}
where $H$ is the SM Higgs doublet, and 
$\mu_H$ is the mass parameter for the Higgs mass term
$\mu_H^2 H^\dag H$.
As we explain at (iv),  the Higgs portal coupling $\lambda_{HS}$ has to be extremely small, and we will ignore it throughout  the following discussions.
So, it  is just introduced to ensure renormalizability.~\footnote{We assume that the $ H^\dag HR$ coupling is negligibly small, such that
the Higgs plays no role in our scenario of inflation.
 }\\
(iv) This part of the model is responsible for making 
the  right- and left-handed
neutrinos, $N_R$ and  $\nu_L$, massive and also
 for generating  the Higgs mass term radiatively:
\al{\frac{{\cal L}_N  }{\sqrt{-g}}&=	 \frac{i}{2} \bar{N}_R \slashed{\partial} N_R 
	- \frac{1}{2} y_M S N^T_R C N_R 	
	-\left( y_\nu \bar{L} \tilde{H} \,\frac{1+\gamma_5}{2}
	N_R + \text{h.c.} \right)\,,
\label{LNR}}\\
where $\tilde{H}=i\sigma_2 H^*$, $L$ is the lepton doublet, and
$C$ is the charge conjugation matrix.
Strictly speaking, the Yukawa couplings $y_\nu$ 
and $y_M$ should be matrices in the generation space. However, we will not consider the flavor structure, instead $y_\nu$ and $y_M$ will be representative real numbers.
The right-handed neutrinos $N_R$ become 
massive due to the second term on the rhs of Eq.~(\ref{LNR})
when  $S$ acquires the VEV, i.e. $m_N = y_M v_S$. 
We assume that $m_N \sim 10^7$ GeV to obtain a desired size of 
the radiative correction 
to $\mu_H^2$ \cite{Vissani:1997ys,Casas:1999cd,Clarke:2015gwa,Bambhaniya:2016rbb}  for triggering the electroweak symmetry breaking
and at the same  to make the type-I seesaw mechanism 
 \cite{Minkowski,Yanagida:1979as,GellMann:1980vs,Goran} 
viable - a scenario dubbed the “Neutrino option” \cite{Brivio:2017dfq, Brivio:2018rzm,Brdar:2018num,Brdar:2018vjq,Brdar:2019iem,Brivio:2019hrj,Aoki:2020mlo,Kubo:2020fdd,Brivio:2020aut}.
In this scenario it is assumed that the  radiative correction 
is  the dominant contribution to $\mu_H^2$
while the tree-level contribution $\lambda_{HS} v_S^2/4$ is 
negligibly small \cite{Brdar:2018vjq,Brdar:2018num}.
\\
\\
As we have seen above,
the real scalar field $S$ appears in all the  sectors.
It is the mediator that transfers the robust energy scale, created
by  the chiral condensate, to the gravity sector, while playing  a role of inflaton
on one hand, and  is on the other hand responsible for generating the
 heavy right-handed neutrinos, which in turn give rise to
 the electroweak symmetry breaking as well as to the light active neutrinos.

 \subsection{Origin of the unified energy scale:\\
 Nambu-Jona-Lasinio description of chiral symmetry breaking }
 
 Lattice gauge theory is a first-principle calculation in
 the QCD-like hidden sector that is  described by the Lagrangian (\ref{LH}).
Here we shall use an effective field theory - the NJL
theory \cite{Nambu:1960xd,Nambu:1961tp,Nambu:1961fr} - to describe
 the dynamical chiral symmetry breaking in the hidden sector.~\footnote{Linear sigma model is used in Ref. \cite{Hur:2011sv}, and 
 the holographic method is applied in Ref.  \cite{Hatanaka:2016rek}.}
Following Refs. \cite{Holthausen:2013ota,Kubo:2014ida,Ametani:2015jla} we assume 
$n_f=n_c=3$, because in this case the meson properties 
in hadron physics can be used to reduce the independent parameters
of the NJL theory
\cite{Holthausen:2013ota,Kubo:2014ida,Ametani:2015jla}.
So, the hidden chiral symmetry is
$\mathrm{SU}(3)_{L}\times\mathrm{SU}(3)_{R}$,
which  is dynamically broken to its diagonal subgroup 
$SU(3)_V$ by the non-zero chiral condensate
$\left<\bar{\psi}\psi\right>$,
implying  the existence of 8 NG bosons.

The  NJL Lagrangian for the hidden sector~\footnote{Here we work in the 
flat space time.}  is given by
 \cite{Nambu:1960xd,Nambu:1961tp,Nambu:1961fr}
 \begin{align}
 {\cal L}_{\rm NJL} &=\mbox{Tr}~\bar{\psi}(i
\slashed{\partial} -{\bm y}S)\psi+2G~\mbox{Tr} ~\Phi^\dag \Phi
+G_D~(\det \Phi+h.c.)\,,
\label{eq:NJL10}
\end{align}
where
\begin{align}
\Phi_{ij}&= \bar{\psi}_i(1-\gamma_5)\psi_j=
\frac{1}{2}\sum_{a=0}^{8}
\lambda_{ji}^a\, [\,\bar{\psi}\lambda^a(1-\gamma_5)\psi\,]\,,
\end{align}
$\lambda^a (a=1,\dots, 8)$ are the Gell-Mann matrices with
$\lambda^0=\sqrt{2/3}~{\bm 1}$, and the canonical dimension of
$G\, (G_D)$ is $-$2 (5).
Further,  we use
the self-consistent mean-field (SCMF) approximation
of Refs. \cite{Kunihiro:1983ej,Hatsuda:1994pi} and define
the mean fields $\sigma_i~(i=1,2,3)$ 
and $\phi_a~(a=0,\dots,8)$  in the ``Bardeen-Cooper-Schrieffer" vacuum as
\begin{align}
\label{varphi}
\sigma_i =- 4 G\left<\bar{\psi}_i \psi_i  \right> \,, ~~~
\phi_a =-2 i G\left<\bar{\psi}_i \gamma_5 \lambda^a\psi_i  \right>\,,
\end{align}
respectively.
Here we suppress the CP-even mean fields corresponding 
to the non-diagonal elements of $\langle \bar{\psi}_i\psi_j \rangle$, because they do not play any role
for our purpose.
\footnote{The lightest of $\sigma_i$ behaves as  the dilaton \cite{Kunihiro:1983ej,Hatsuda:1994pi}, the (quasi) NG boson associated
with the spontaneous breaking of scale invariance. It is massive, because
scale invariance is explicitly broken by anomaly at the fundamental level
and by four and six Fermi interactions (\ref{eq:NJL10}) at the effective level. }
Then splitting the NJL Lagrangian $\mathcal{L}_{\text{NJL}}$ into two parts as 
 $\mathcal{L}_{\text{NJL}} =\mathcal{L}_{\text{MFA}}+\mathcal{L}_{I}$ 
 where $\mathcal{L}_{I}$ is normal ordered (i.e., $\langle 0|\mathcal{L}_{I}|0\rangle =0$),
 we find the mean-field Lagrangian $\mathcal{L}_{\text{MFA}}$ in the SCMF approximation:  
 ~\footnote{The mean-field Lagrangian $\mathcal{L}_{\text{MFA}}$
 in the case of broken $SU(3)_V$ can be found in Ref. \cite{Ametani:2015jla}.}
\begin{align}
\nonumber
   \mathcal{L}_{\text{MFA}}= &   
   \mathrm{Tr} ~\bar{\psi}(i\Slash{\partial}-M)\psi -i\mathrm{Tr} ~\bar{\psi}\gamma_5 \phi \psi -\frac{1}{8G}\left( 3\sigma^2+2\sum^8 _{a=1} \phi_a \phi_a \right)  \\
 \label{Hidden SCMFA}
    & +\frac{G_D}{8G^2}\left(  -\mathrm{Tr} ~\bar{\psi} \phi^2 \psi + \sum^8 _{a=1} \phi_a \phi_a \mathrm{Tr} ~\bar{\psi}\psi + i\sigma \mathrm{Tr}~ \bar{\psi}\gamma_5 \phi \psi +\frac{\sigma^3}{2G}+\frac{\sigma}{2G}\sum^8 _{a=1} (\phi_a)^2   \right)\,, 
\end{align}
where $\phi=\sum_{a=1}^8~\phi_a \lambda^a$, $\sigma=\sigma_1=
\sigma_2=\sigma_3$, and
$\phi_0$ has been suppressed.
The constituent fermion mass $M$ is given by 
\begin{align}
\label{M}
M(S,\sigma)= \sigma+yS-\frac{G_D}{8G^2}\sigma^2\,,
\mbox{where}~y=y_1=y_2=y_3\,.
\end{align}

The one-loop effective potential can be obtained 
from 
$\mathcal{L}_{\text{MFA}}$
(\ref{Hidden SCMFA}) by integrating out the hidden fermions:
\begin{align}
V_{\rm NJL}(S,\sigma)
& = \frac{3}{8G}\sigma^2-
\frac{G_D}{16G^3}\sigma^3
-3n_c I_0(M,\Lambda_H)\,,
\label{eq:Vnjl}
\end{align}
where the  function $I_0$ is given by
\begin{align}
  I_0(M, \Lambda)= \frac{1}{16\pi^2}\left[ \Lambda^4 \ln \left( 1+\frac{M^2}{\Lambda^2 }\right)-M^4 \ln \left( 1+\frac{\Lambda^2 }{M ^2}\right) + \Lambda^2 M ^2\right] \,,
\end{align}
with a four-dimensional momentum cutoff $\Lambda$. 
\footnote{
The one-loop correction $I_0(M,\Lambda)$
 in a curved space time has been calculated in Ref. 
\cite{Inagaki:1993ya,Inagaki:1997kz}. It is proportional to
$(\sigma^2/96 \pi^2)\,R$, which is a minimal coupling like
 $\beta\,S^2R$.  However, since $\beta=O(1) - O(10^4)$
as we will see when discussing inflation, 
this additional term in the curved space time is negligible compared
to the tree-level term $\beta\,S^2R$. }
Note that the cutoff parameter $\Lambda$ is an additional free parameter in the NJL theory.
For a certain interval of  the dimensionless 
parameters, $G^{1/2}\Lambda$ and $(-G_D)^{1/5}\Lambda$,
we have $v_\sigma=\left<\sigma\right>\ne 0$
and $v_S=\left<S \right>\ne 0$
\cite{Holthausen:2013ota,Kubo:2014ida,Ametani:2015jla}.
The actual value  of $\Lambda$ can be fixed, once the hidden sector is connected
with a sector whose energy scale is given.
In our case, the hidden sector is coupled via the mediator $S$ 
with the gravity sector (ii) described by
Eq.~(\ref{LCGR}) as well as  with the right-handed neutrino sector (iv) 
described by Eq.~(\ref{LNR}),
while the coupling with the SM sector (iii)
 is assumed to be extremely suppressed,
because we assume that the portal coupling $\lambda_{HS}$ is negligibly small.
Hereafter we will denote  the cutoff  in our hidden sector  by $\Lambda_H$.
It can be fixed in the following way.

The NJL parameters for the SM  hadrons satisfy
the dimensionless relations
\cite{Holthausen:2013ota,Kubo:2014ida,Ametani:2015jla}
$ \left.G^{1/2}\Lambda\,\right|_\text{Hadron}=1.82~\mbox{and}~
~~\left.(-G_D)^{1/5}\Lambda\,\right|_\text{Hadron}=2.29$\,,
where $\Lambda_\text{Hadron} \simeq 0.1$ GeV.
We assume that  
 the above dimensionless relations are satisfied
 while scaling-up the values of
$G, G_D$ and the cutoff $\Lambda$ from QCD hadron physics:
\begin{align}
 \left. G^{1/2}\Lambda  \,\right|_\text{Hidden} &=1.82\,,
 ~~~~\left.(-G_D)^{1/5}\Lambda\,\right|_\text{Hidden}=2.29\,,
 \label{NJL para}
\end{align}
 should remain unchanged for $\Lambda_\text{Hidden}=\Lambda_H
 \gg \Lambda_\text{Hadron}$.

It is noted that the mean fields $\sigma$ and $\phi_a$ are non-propagating classical fields at the tree level.
Therefore, their kinetic terms are generated by integrating out the hidden fermions at the one-loop level, which will be seen in Section \ref{Mass spectrum}, where 
two point functions are calculated.
Further, one can see that 
the potential $V_{\rm NJL}(S,\sigma)$ is asymmetric in $\sigma$ by inspecting the last term in the NJL Lagrangian in Eq.~(\ref{Hidden SCMFA}) as well as the constituent mass $M$ in Eq.~(\ref{M}); due to latter chiral phase transition can become of first-order 
\cite{Aoki:2017aws,Helmboldt:2019pan,Aoki:2019mlt}.

 \subsection{Planck mass}
We next integrate out the quantum fluctuations $\delta S$  at one-loop 
to obtain the effective potential
  \al{
  	U_S(S,R)&=   \frac{1}{4}\lambda_S S^4
  	+\frac{1}{64 \pi^2}\left(\,
  	\tilde{m}_s^4 \ln  [\tilde{m}_s^2/\mu^2]
  	\,\right)\,,
  	\label{Ueff}}
  where
  \al{\label{ms_msigma}
	\tilde{m}_s^2 &=
  	3 \lambda_S S^2+\beta R\,.
  }
  Here we have used the $\overline{\mbox{MS}}$ scheme, and
  the constant $-3/2$ is absorbed into the renormalization scale $\mu$.~\footnote{
  Strictly speaking,  $\beta$ in Eq.~\eqref{ms_msigma} should read $\beta-1/6$, 
  if one properly takes into account the non-flatness of space-time background and the integration of the quantum fluctuations \cite{Markkanen:2018bfx}. 
  However, since $\beta$ will turn out be large (i.e.$\gsim 4$) for realistic cosmic inflation, we will be ignoring the constants $1/6$ throughout the paper.
  }
Our total effective potential in the Jordan frame now reads 
\al{
U_\mathrm{eff}(S,\sigma,R) &= 
V_\text{NJL}(S,\sigma) +U_S(S,R)-U_0\,,
\label{Ueff2}
}
where $V_\text{NJL}$ is given in Eq.~(\ref{eq:Vnjl}), and $U_0$ is the zero-point energy density.
We have subtracted it, such that
$U_\mathrm{eff}(S=v_S,\sigma=v_\sigma,R=0)=0$ is satisfied.
The zero-point energy density $U_0$ is negative, because it is a consequence of the spontaneous breaking of global conformal symmetry. 
Subtracting $U_0$ means that we start with a nonzero
cosmological constant. That is, 
we add an explicit super-soft breaking of scale invariance at tree level,
and accordingly   we put  the cosmological constant problem
\cite{Weinberg:1988cp}  aside here and continue with our discussion.

  To compute  $v_S=\langle S\rangle$ and 
   $v_\sigma=\langle \sigma\rangle$, 
   we assume  that $\beta  R  < 3 \lambda_S S^2 $ (during inflation), such that
$U_S(S,R)$ in  Eq.~(\ref{Ueff}) can be expanded 
  in powers of 
  $\beta R$:
  \al{
  	U_S(S,R) &=
  	U_\mathrm{CW}(S)+ U_{(1)}(S) \,R+U_{(2)}(S) R^2 +O(R^3 )\,,
  	\label{Ueff1}
  }
  where 
  \al{  
  	U_\mathrm{CW}(S) &=\frac{1}{4}\lambda_S S^4+
	\frac{9}{64 \pi^2}
  	 \, \lambda_S^2 S^4\,\left(\,
	 \ln [\,3 \lambda_S S^2/v_S^2\,]-\frac{1}{2}\,\right)\,,\label{Ucw}\\
  	U_{(1)}(S) &=\frac{3}{32 \pi^2}
  	\beta  \lambda_S S^2\,\ln [\,3 \lambda_S S^2/v_S^2\,]\,,
  	\label{U1}\\
  	U_{(2)}(S) & =\frac{1}{64 \pi^2}
  \beta^2  \left(1+\ln [\,3 \lambda_S S^2/v_S^2\,]\right)\,.
  	\label{U2}
  } 
We have chosen $\mu^2=v_S^2/2$, such that
we can calculate $v_S$ from $V_\text{NJL}(S,\sigma)+\lambda_S S^4/4$
(i.e. the logarithmic term does not enter into the  determination of $v_S$
and $v_\sigma$.).
Since we are assuming a negligibly small (but, of course, nonzero during inflation) value of the curvature scalar $R$, 
we obtain the $R$-independent leading-order $v_S$ and $v_\sigma$ from $V_\text{NJL}(S,\sigma)+\lambda_S S^4/4$.
Finally,  the  identification of $M_\mathrm{Pl}$ follows from 
  the first term in Eq.~(\ref{LCGR}) along with Eq.~(\ref{Ueff1}): 
  \al{
  	M_\mathrm{Pl} &= v_S \left( \beta +
  	\frac{2 U_{(1)}(v_S)}{v_S^2} \right)^{1/2} =
	\sqrt{\beta}\, v_S\left(
 1+\frac{3\lambda_S}{16\pi^2}\ln[3\lambda_S]\right)^{1/2}\,.
  	\label{mpl}
  }
 As we will see, $\beta$ is of order $10^3$ for a successful inflation,
 $v_S$ is few orders of magnitude smaller than $M_\mathrm{Pl}$.

   \subsection{Neutrino mass and Higgs mass}
   The basic idea of the neutrino option is already explained in Section
   \ref{Model}.
 Here we  briefly discuss how the mass hierarchy is realized in our model.
   Since for the neutrino option to work, we require  $m_N=y_M v_S\sim 10^7$
   GeV, where $y_M$ is the Majorana-Yukawa coupling
   in the sector (iv) described  by the Lagrangian (\ref{LNR}).
   On the other hand, we have $v_S\simeq M_\mathrm{Pl}/\sqrt{\beta}$
   from Eq.~(\ref{mpl}),
   which implies $y_M\sim \sqrt{\beta}\, 10^{-11}$.
Therefore, the Majorana-Yukawa coupling 
$y_M$ has to be  very small. Note however that the small $y_M$ is not unnatural,
because in its absence the lepton number is conserved.
Compared with $m_H/M_\mathrm{Pl}\sim 10^{-16}$,
where $m_H\simeq 125$ GeV is the SM Higgs mass,
this original mass hierarchy $10^{-16}$ can be largely  softened 
for $\beta$ large.
Note also that $m_H^2\sim y_\nu^2 m_N^2/4\pi^2$ in the neutrino option,
where $y_\nu$ is the Dirac-Yukawa coupling.
This equation implies $m_H^4\sim \lambda_H m_\nu m_N^3/4\pi^2$,
where  $\lambda_H$  is the Higgs quartic coupling in Eq.~(\ref{LSM})
and $m_\nu$ stands for  the light neutrino mass. That is, the scale
of the Higgs mass is basically fixed by the neutrino sector,
which is a consequence of the fact that the SM is only indirectly coupled
with the hidden sector through a negligibly small portal coupling $\lambda_{HS}$.

\section{Inflation}\label{sec3}
\subsection{Effective action for inflation}

If the inequality $\beta  R  < 3 \lambda_S S^2 $ is satisfied during inflation,
the higher order terms in Eq.~(\ref{Ueff1}) 
can be  consistently neglected for inflation. We will proceed with this 
simplification, but we will posteriori check whether the inequality
is satisfied.
Similarly, if $\kappa$, the coefficient
of the $W_{\mu\nu\alpha\beta}W^{\mu\nu\alpha\beta}$ term
in Eq.~(\ref{LCGR}), is small, this term  has only a small effect on the 
inflationary parameters (see for instance \cite{Ghilencea:2019rqj}),
so that we will ignore it in the following discussion as well.
\footnote{The presence of the $W_{\alpha\beta\mu\nu} W^{\alpha\beta\mu\nu} $ 
in (22) causes
 theoretical problems:
The classical Hamiltonian is bounded from below (Ostrogradsky instability).
Furthermore,  the massive 
(after the spontaneous scale symmetry breaking) spin two state 
is a ghost state \cite{Stelle:1977ry}, due to the wrong sign of its
kinetic term (see also \cite{Alvarez-Gaume:2015rwa}). Therefore, 
this state endangers unitarity of the theory. 
Although  no ultimate solution seems to exist
 to this problem at present, there are
various interesting Ans\"atze, which is reviewed for instance in \cite{Salvio:2018crh}.
We do not address this problem here, because 
this would be beyond the scope of the present paper.}
In doing so, we arrive at
the effective Lagrangian  for inflation in the Jordan frame
 \setcounter{equation}{0}
\renewcommand\theequation{3.\arabic{equation}}
\al{
	\frac{{\cal L}_\mathrm{eff}}{\sqrt{-g_J}} 
	=-\frac{1}{2}M_\mathrm{Pl}^2 B(S) R_J+
	G(S) R_J^2+\frac{1}{2}g_J^{\mu\nu}\left(\partial_\mu S\partial_\nu S+
	Z_\sigma^{-1}(S,\sigma) \partial_\mu \sigma\partial_\nu \sigma
\right)
	-U(S,\sigma)\,,
	\label{Leff}
}
where
\al{
  B(S) =&
\frac{\beta\, S^2}{M_\text{Pl}^2}\left(
 1+\frac{3\lambda_S}{16\pi^2}
 \ln[3\lambda_S S^2/v_S^2]\right)\,,
 \label{BoS}\\
  G(S) =&\gamma
-\frac{\beta^2}{64\pi^2}\left(
 1+
 \ln[3\lambda_S S^2/v_S^2]\right)\,,
 \label{GoS}\\
 U(S,\sigma) =&V_\text{NJL}(S,\sigma)+\frac{\lambda_S}{4} S^4+
 \frac{9\lambda_S^2 S^4}{64\pi^2}\left(-\frac{1}{2}+
 \ln[3\lambda_S S^2/v_S^2]\right)-U_0\,.\label{UoS}
 }
Here $g_J^{\mu\nu}\, (g_J=\det g_{\mu\nu}^J)$ and $R_J$ denote the inverse of
the metric $g_{\mu\nu}^J$ and  the Ricci scalar of Jordan-frame space time, respectively.
To remove the $R_J^2$ term from Eq.~(\ref{Leff}), we
introduce an auxiliary field $\chi$ with mass dimension two
and replace 
$ G(S) R_J^2$ by 
$2  G(S) R_J\chi-G(S) \chi^2$.
Then performing a Weyl rescaling of the metric,
$g_{\mu\nu} = \Omega^2\, g_{\mu\nu}^J$ with
\begin{align}
\label{eq:WeylTrafo}
\Omega^2(S,\chi) =
B(S) - \frac{4\,G(S)\chi}{M_{\rm Pl}^2}\,,
\end{align}
we arrive at the Einstein frame with the Lagrangian
\begin{align}
\label{eq:LagEO}
\frac{\mathcal{L}_{\rm eff}^E}{\sqrt{- g}} = &-\frac{1}{2}\,M_{\rm Pl}^2
\left(R - \frac{3}{2}\,g^{\mu\nu}\,\partial_\mu \ln\Omega^2(S,\chi)\,
\partial_\nu \ln\Omega^2(S,\chi)\right)\nn\\
& + \frac{g^{\mu\nu}}{2\,\Omega^2(S,\chi)}\left(
\,\partial_\mu S\,\partial_\nu S+Z_\sigma^{-1}(S,\sigma)\partial_\mu \sigma\,\partial_\nu \sigma
\right)
-V(S,\sigma,\chi) \,,
\end{align}
where 
\begin{align}
V(S,\sigma,\chi)  =
\frac{U(S,\sigma) + G(S) \chi^2}
{\left[\,B(S) M_{\rm Pl}^2 -
	4\,G(S)\chi\,\right]^2} \, M_{\rm Pl}^4 \,.
\label{VSpsi}
\end{align}
Note that $\chi$ is promoted to a propagating scalar field in the Einstein frame.
Using the scalaron field $\varphi$~\cite{Barrow:1988xh,Maeda:1988ab}, 
which is canonically normalized and  defined as
\begin{align}
\label{eq:phi}
\varphi = \sqrt{\frac{3}{2}}\,M_{\rm Pl} \ln\left|\Omega^2\right| \,,
\end{align}
we finally obtain the Einstein-frame Lagrangian for the coupled $S$-$\sigma$-scalaron system:
\begin{align}
\label{eq:LEphichi}
\frac{\mathcal{L}_{\rm eff}^E}{\sqrt{- g}} =& -\frac{1}{2}\,M_{\rm Pl}^2\,R
+ \frac{1}{2}\,g^{\mu\nu}\,\partial_\mu\varphi\,\partial_\nu \varphi 
+ \frac{1}{2}\,e^{-\Phi(\varphi)}\,g^{\mu\nu}\left(\,\partial_\mu
S\,\partial_\nu S+Z_\sigma^{-1}(S,\sigma)\partial_\mu \sigma\,\partial_\nu \sigma
\right)\nn\\
& - V(S,\sigma,\varphi)  \,,
\end{align}
where $\Phi\left(\varphi\right) = \sqrt{2/3}\,\varphi/M_{\rm Pl}$, and
the potential $V$ given in Eq.~(\ref{VSpsi}) 
is
\begin{align}
V(S,\sigma,\varphi) = 
e^{-2\,\Phi(\varphi)} \left[ U(S,\sigma)
+ \frac{M_{\rm Pl}^4}{16\,G(S)}\left(B(S)
- e^{\Phi(\varphi)}\right)^2\right] \,.
\label{VSphi}
\end{align}

\subsection{Valley approximation}
\begin{figure}[ht]
\begin{center}
\hspace{0.7cm}
\includegraphics[width=2.5in]{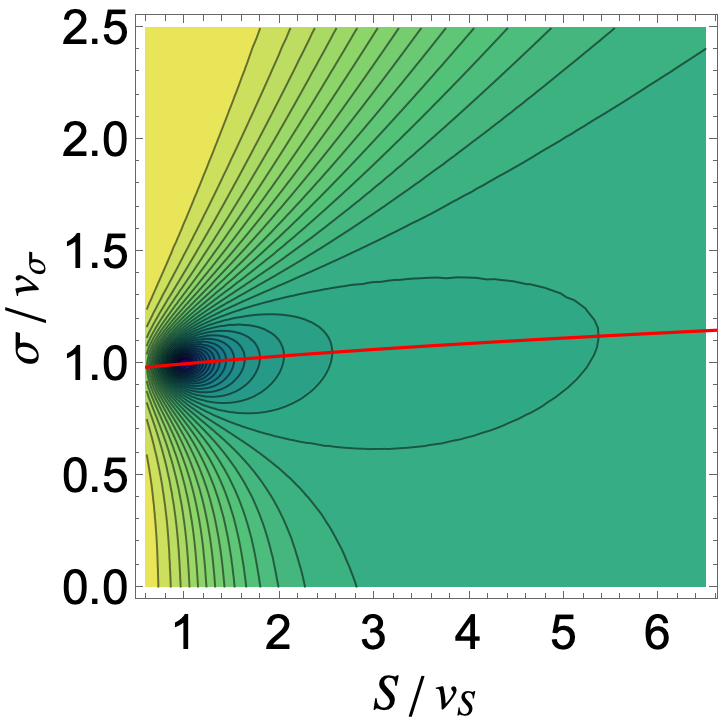}\hspace{0.9cm}
\includegraphics[width=2.78in]{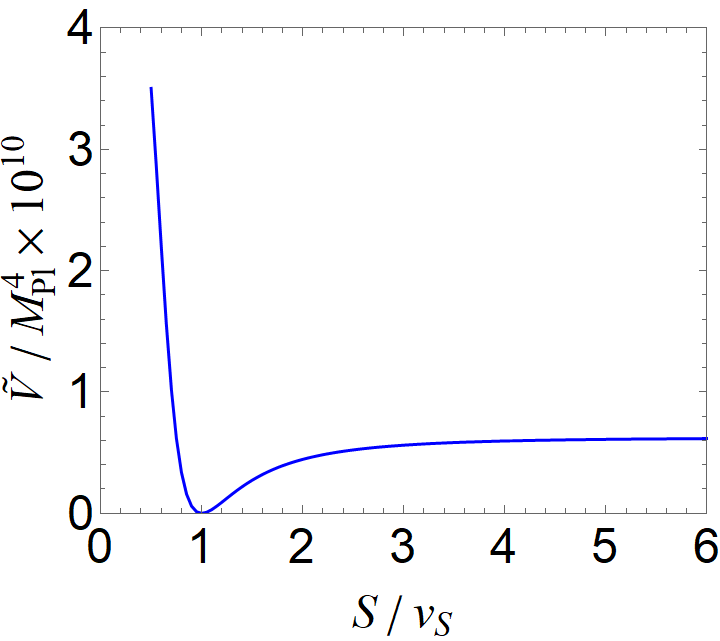}
\caption{Left: The contour plot of $\tilde{V}(S,\sigma)$, where
$\tilde{V}(S,\sigma)=V(S,\sigma,\varphi_\text{v})$. The red line is
the bottom line of  the valley $\tilde{V}(S,\sigma)$.
Right: $\tilde{V}(S,\sigma)/M_\text{Pl}^4$ along the bottom line of the valley
(the red line of the left panel) against $S/v_S$.
}
\label{Vtilde-contour}
\end{center}
\end{figure}

\begin{figure}[ht]
\begin{center}
\includegraphics[width=3in]{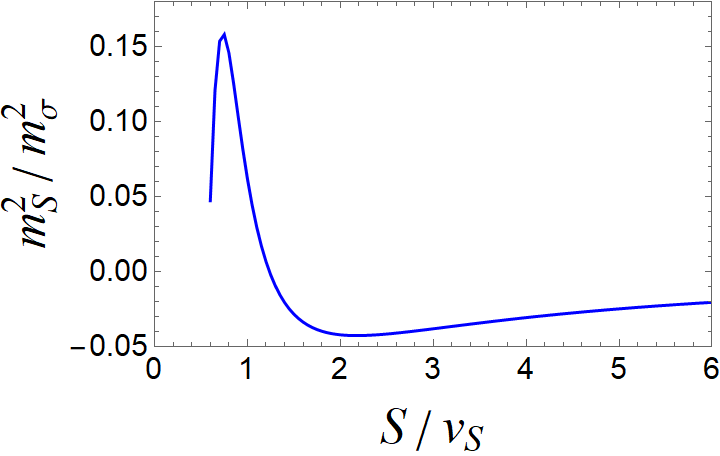}
\caption{
$m_S^2/m_\sigma^2$ against $S/v_S$ along the bottom line of the valley.
The graph shows that the perpendicular direction to the bottom line of the valley  
is much steeper than the parallel direction, so that the inflaton can roll down
slowly along the  bottom line of the valley.
}
\label{mass-ratio}
\end{center}
\end{figure}

As we see from the Lagrangian (\ref{eq:LEphichi}) 
we have a multi-field system at hand \cite{Wands:2007bd}.
Fortunately, it turns out that the valley approximation \cite{Kannike:2015apa}
can be successfully applied, so that we only have to deal with a single-field inflaton 
system, as we will see below.
To begin with we find that the stationary point condition
\al{
\left.\frac{\partial V(S,\sigma,\varphi)}{\partial \varphi}\,\right|_{ \varphi= \varphi_\text{v}} =0
}
can be solved analytically for $\varphi$:
\al{
\varphi_\text{v}=&\sqrt{3/2}M_\text{Pl}\, \ln
[\,B(S)+4 A(S,\sigma) B(S)\,]\,,~\mbox{where}~ A(S,\sigma)=\frac{4 G(S) U(S,\sigma)}{B^2(S)M_\text{Pl}^4}\,.
\label{phiv}
}
Therefore, the three-field system potential $V(S,\sigma,\varphi)$
can be reduced to a double-field system potential 
$\tilde{V}(S,\sigma)=V(S,\sigma,\varphi_\text{v})$, which takes the following form: 
\al{
	\tilde{V}(S,\sigma) &
	= \frac{U(S,\sigma) 
		M_\text{Pl}^4}{\,16 G(S) \,U(S,\sigma) + B^2(S)\, M_\text{Pl}^4} \,.\label{Vtilde}
}
In Fig. \ref{Vtilde-contour} (left) we show a contour plot of $\tilde{V}(S,\sigma)$ 
for
\al{
y=4.00\times 10^{-3}\,,\quad \lambda_S=1.14\times 10^{-2}\,,\quad
\beta =6.31\times 10^3\,,\quad \gamma =1.26\times 10^8\,.
\label{bench1}
}
The red line in Fig. \ref{Vtilde-contour} (left)  is the bottom line of the valley,
and 
we will assume that the inflaton slowly rolls down along this line.
$\tilde{V}(S,\sigma)/M_\text{Pl}^4$ along the 
bottom line of the valley  against $S/v_S$ is plotted in Fig.  \ref{Vtilde-contour}
(right), from which we see that the potential $\tilde{V}(S,\sigma)$
for $S/v_S >1$ along this line is very flat.
  In Fig. \ref{mass-ratio} we plot
 $m_S^2/m_\sigma^2$ against $S/v_S$ along the  bottom line, where
 $m_S^2=\partial^2 \tilde{V}(S,\sigma)/\partial S^2$ and
  $m_\sigma^2=\partial^2 \tilde{V}(S,\sigma)/\partial \sigma^2$.
We see from Fig. \ref{mass-ratio} that the second derivative  of 
$\tilde{V}(S,\sigma)$ 
with respect to $\sigma$   is much larger than that
 with respect to $S$ along the bottom line, meaning that
the perpendicular direction to the bottom line of the valley  
is much steeper than the parallel direction. This justifies the assumption above
that the inflaton slowly rolls down along the bottom line of the valley.

\begin{figure}[ht]
\begin{center}
\hspace{0.7cm}
\includegraphics[width=2.6in]{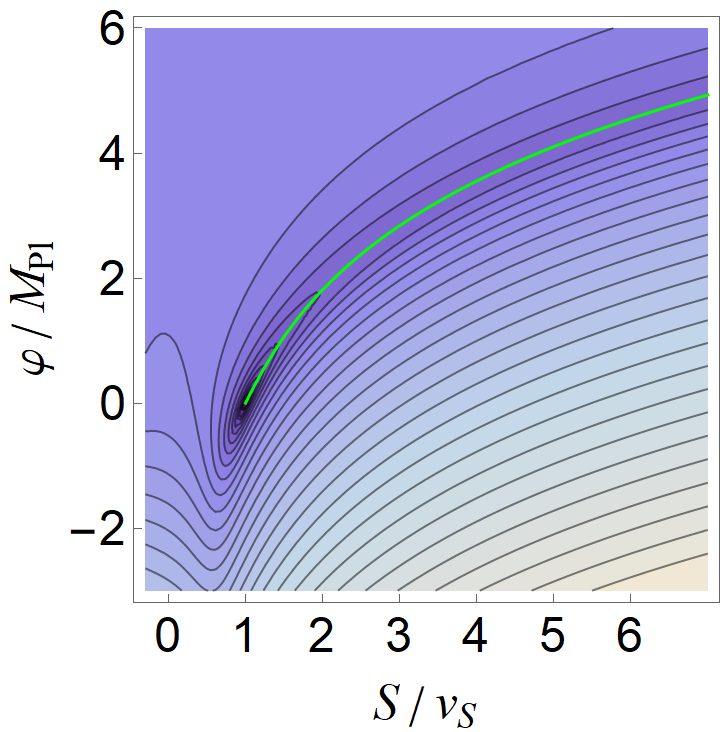}\hspace{0.5cm}
\includegraphics[width=2.78in]{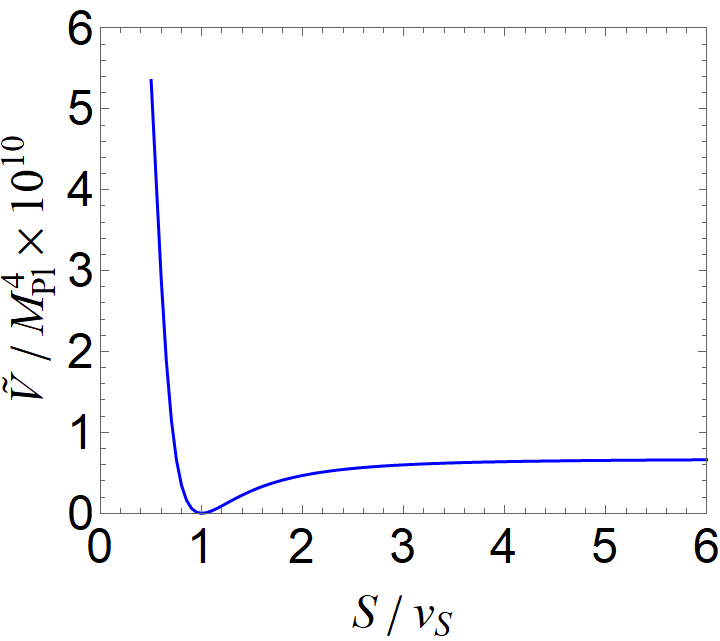}
\caption{Left: The contour plot of $V(S,\sigma_\text{v}(S),\varphi)$,
where $\sigma_\text{v}(S) $ is the bottom line of  $\tilde{V}(S,\sigma)$
(the red line of Fig.~(\ref{Vtilde-contour}),
 where of  $V(S,\sigma,\varphi)$ 
is given in Eq.~(\ref{VSphi}).
The green line is the bottom line of
 $V(S,\sigma_\text{v}(S),\varphi)$, along which the inflaton slowly rolls down.
 Right: $V(S,\sigma_\text{v}(S),\varphi)/M_\text{Pl}^4$ against $S/v_S$
 along the bottom line of  $V(S,\sigma_\text{v}(S),\varphi)$
 (the blue line of the left panel).
}
\label{V-S-phi}
\end{center}
\end{figure}

We next consider the potential $V(S,\sigma_\text{v}(S),\varphi)$ given  in Eq.~(\ref{VSphi}),
where $\sigma_\text{v}(S)$
is the bottom line of $\tilde{V}(S,\sigma)$.
Its contour plot is shown in Fig. \ref{V-S-phi} (left) for the same set of the parameters 
as given in Eq.~(\ref{bench1}).
We see from the left panel that the potential 
$V(S,\sigma_\text{v}(S),\varphi)$ has a desired valley structure.
We also see from the right panel that 
the potential $V(S,\sigma_\text{v}(S),\varphi)$ for $S/v_S > 1$ along the 
bottom line (the green line of the left panel) is very flat.

\subsection{Prediction of the inflationary parameters}
The effective Lagrangian for the single-field inflaton system can be obtained from the Lagrangian~(\ref{eq:LEphichi}), where we treat $S$ as the independent field variable
for the single-field inflaton system. Since $\sigma$ and $\varphi$ in this case are functions of $S$, their kinetic terms become a part of 
the kinetic term for $S$:
\al{
& e^{-\Phi(\varphi_\text{v}(S))}\,g^{\mu\nu}\left[\,
	\partial_\mu \,S \partial_\nu\, S +	Z_\sigma^{-1} (S,\sigma_\text{v})
	\partial_\mu \sigma_\text{v}(S)\, \partial_\nu\, \sigma_\text{v}(S) \right]
 +
	g^{\mu\nu}\partial_\mu \, \varphi_\text{v}(S) \partial_\nu \,\varphi_\text{v}(S)
	\nn\\
	& =F(S)^2 g^{\mu\nu}\partial_\mu \,S \partial_\nu \,S\,, 
	\label{FS}
}
where
\begin{align}
F(S) &= \frac{1}{\left[1+4 \, A(S,\sigma_\text{v}(S) ) \right] B(S)}
\Large\left\{\left[1+Z_\sigma^{-1} (S,\sigma_\text{v}(S) ) (\sigma'_\text{v}(S) )^2
\right]\,\left[1+4\,A(S,\sigma_\text{v}(S) )\right]B(S)
\right.\nn\\
&\left.+ \frac{3}{2}\,M_{\rm Pl}^2
\big\{\left[1+4\,A(S,\sigma_\text{v}(S) )\right]B'(S)
+ 4\,A'(S,\sigma_\text{v}(S) )B(S)\big\}^2\right\}^{1/2}\,,
\label{field_norm}
\end{align}
with $A(S,\sigma)$ and $B(S)$ given in Eqs. (\ref{phiv}) and (\ref{BoS}), respectively, and the prime stands for derivative with respect to $S$.
We finally arrive at the single-field inflaton system, which is described by
\begin{align}
\frac{\mathcal{L}_{\rm eff}^E}{\sqrt{- g}} =
- \frac{1}{2}\,M_{\rm Pl}^2\,R
+ \frac{1}{2}\,F(S)^2\,g^{\mu\nu}
\,\partial_\mu S\,\partial_\nu S
- V_{\rm inf}(S) \,,
\label{LS}
\end{align}
with 
\al{
V_{\rm inf}(S) &=V(S,\sigma_\text{v}(S),\varphi_\text{v}(S))\,,
\label{Vinf}
}
where $V(S,\sigma,\varphi)$ and $\varphi_\text{v}$ are given in Eqs. (\ref{VSphi}) and (\ref{phiv}), respectively, and
$\sigma_\text{v}$ is the bottom line of $\tilde{V}(S,\sigma)=V(S,\sigma,\varphi_\text{v})$.
Note that the canonically normalized inflaton field $\hat{S}$ 
can be obtained from
\begin{align}
\hat{S}(S) = 
\int_{v_S}^S dx \,F(x) \,.
\label{Shat}
\end{align}
However, to compute the slow roll parameters we will use $S$ instead of ${\hat S}$:
\begin{align}
\label{epsilon}
\varepsilon (S) & = 
\frac{M_{\rm Pl}^2}{2\,F^2(S)}
\left(\frac{V'_{\rm inf}
(S)}{V_{\rm inf}(S)}\right)^2 \,,
\\ 
\eta(S) & =
\frac{M_{\rm Pl}^2}{F^2(S)} \left(
\frac{
V''_{\rm inf}(S)}{V_{\rm inf}(S)} -
\frac{ F'(S)}{F(S)}
\frac{V'_{\rm inf}(S)}{V_{\rm inf}(S)}\right) \,.
\label{eta}
\end{align}

The number of e-folds $N_e$ can be computed from 
\begin{align}
N_e =\int_{S_\mathrm{end}}^{S_*}dS \;
\frac{F^2(S)}{M_{\rm Pl}^2}
\frac{V_{\rm inf}(S)} {V'_{\rm inf}(S)} \,,
\label{efolding}
\end{align}
where ${S_*}$ is the value of
$S$ at the time of CMB horizon exit, and $S_\mathrm{end}$  is 
that of $S$ at the end of
inflation, i.\e.\ $\varepsilon (S=S_\mathrm{end}) =1$.
The CMB observables - the scalar power spectrum amplitude
$A_s$, the scalar spectral index $n_s$ and
the tensor-to-scalar ratio $r$ -
can be obtained from
\begin{align}
A_s = \frac{V_{\rm inf\,*}}{24\pi^2\,\varepsilon_*\,M_{\rm Pl}^4} \,, \quad
n_s = 1 + 2\,\eta_* - 6\,\varepsilon_* \,, \quad
r = 16\,\varepsilon_* \,,
\label{parameters}
\end{align}
where the quantities with $*$ are evaluated at $S=S_*$.
In the following discussions we constrain the parameter space spanned by
$y, \lambda_S,
\beta$ and $\gamma$, such that 
\al{
\ln (10^{10} A_s) = 3.044 \pm 0.014~\mbox{and}~
50 \lsim N_e \lsim 60
\label{As_constraint}
}
are satisfied \cite{Aghanim:2018eyx,Akrami:2018odb}.

For the set of the parameters (\ref{bench1}) we obtain 
\al{
n_s=0.964\,,\quad r=2.00\times 10^{-3}\,,\quad\ln (10^{10} A_s)=3.04\,,\quad N_e=55.5\,,\label{bench1-1}
}
where
\al{v_\sigma/M_\text{Pl} &=8.86\times 10^{-3}\,,\quad
v_S/M_\text{Pl}=1.26\times 10^{-2}\,,\quad
U_0/M_\text{Pl}^4=-7.02\times 10^{-10}\,,\nn\\
S_\text{end}/v_S &=1.40\,,\quad
S_*/v_S=6.25\,,\quad
\Lambda_H/M_\text{Pl} =5.33\times 10^{-2}\,,
\label{bench1-2}
}
and for 
\al{
y& =4.00\times 10^{-4} \,,\quad
\lambda_S=1.2\times 10^{-2} \,,\quad
\beta=1.435\times
10^3 \,,\quad \gamma=5.293\times 10^8\,,
}
we obtain 
\al{
n_s& =0.963\,,\quad r=3.44\times 10^{-3} \,,\quad\ln (10^{10} A_s)=3.04 \,,\quad  N_e=55.0\,,\nn\\
v_\sigma/M_\text{Pl} &=3.97\times 10^{-2}\,,\quad
v_S/M_\text{Pl}=2.64\times 10^{-2}\,,\quad U_0/M_\text{Pl}^4=-2.36\times 10^{-7}\,,\nn\\
S_\text{end}/v_S &=1.09\,,\quad
S_*/v_S=2.30\,,\quad
\Lambda_H/M_\text{Pl} =2.48\times 10^{-1}\,.
\label{bench2-1}
}

\subsection{Numerical study}
\subsubsection{Independent parameters}
Using  the method described in the previous sections, we now
scan the parameter space spanned by  $\lambda_S,y,\beta,\gamma$.
As we will show below, the inflational parameters, $n_s, r$
and $\beta^2A_s$, depend approximately only
on $\lambda_S,y,\bar{\gamma}$,
where 
\al{\bar{\gamma}&=\frac{\gamma}{\beta^2}\,,
\label{gammahat}
}
which makes a  comprehensive analysis easier.
To see this, we recall how the final 
single-field potential $V_\text{inf}(S)=V(S,\sigma_\text{v},\varphi_\text{v})$ (\ref{Vinf}) is 
obtained, where $V(S,\sigma,\varphi)$ and $\varphi_\text{v}$ are given 
in Eqs. (\ref{VSphi}) and (\ref{phiv}), respectively,
while $\sigma_\text{v}$ should be calculated from
$\left.\partial \tilde{V}(S,\sigma)/\partial \sigma\right|_{\sigma=\sigma_\text{v}}=0$,
where $\tilde{V}(S,\sigma) = V(S,\sigma,\varphi_\text{v})$ \eqref{Vtilde}. 
As we see from Eq.~(\ref{UoS}), only  $U(S,\sigma)$ depends on $\sigma$,
where the $\sigma$ dependence enters due to the NJL potential
$V_\text{NJL}(S,\sigma)$ (\ref{eq:Vnjl}). Therefore, we find
\begin{align}
0&=\left.\frac{\partial \tilde{V}}{\partial\sigma}\right|_{\sigma=\sigma_\text{v}}
=  
\left.\frac{B^2 M_\text{Pl}^8 }{(16 GU+ B^2 M_\text{Pl}^4)^2 }  \frac{\partial U}{\partial\sigma}\right|_{\sigma=\sigma_\text{v}}
\to 0=\left.
\frac{\partial V_\text{NJL}}{\partial\sigma}\right|_{\sigma=\sigma_\text{v}} \, .
\label{parsigma}
\end{align}
Note that  $V_\text{NJL}(S,\sigma)$ does not depend
on $\lambda_S, \beta$ and $\gamma$, which implies that 
$\sigma_\text{v}$ does not depend on $\lambda_S, \beta$ and $\gamma$.

As a next step, we redefine $B(S), G(S)$ and $\tilde{V}(S,\sigma)$ as follows:
 \begin{align}
 \bar{B}(S) &= \frac{B(S) M_\text{Pl}^2}{\beta}
 = S^2\left(
 1+\frac{3\lambda_S}{16\pi^2}
 \ln[3\lambda_S S^2/v_S^2]\right)\,,  \\
 \bar{G}(S) &= \frac{G(S)}{\beta^2}
 = \bar{\gamma} -  \frac{ 1+
 	\ln[3\lambda_S S^2/v_S^2] }{64\pi^2} \,,  \\
 \bar{V}(S,\sigma) &= \frac{\beta^2}{M_\text{Pl}^4} \tilde{V}(S,\sigma) = \frac{\beta^2U(S,\sigma)}{16 G(S) U(S,\sigma) + B(S)^2 M_\text{Pl}^4}
 = \frac{U(S,\sigma)}{16 \bar{G}(S) U(S,\sigma) + \bar{B}(S)^2 } \,.
 \label{GbarandBbar}
 \end{align} 
Thus, the $\beta$ dependence disappears in the above functions,
if one uses $\bar{\gamma}$ (\ref{gammahat}) as an independent parameter.
Recalling $V_\text{inf}(S)=\tilde{V}(S,\sigma_\text{v})$,
we find 
that
$V'_\text{inf}/V_\text{inf}$ and  $V''_\text{inf}/V_\text{inf}$,
which enter in $\varepsilon$ (\ref{epsilon}), $\eta$ (\ref{eta}) and 
$N_e$ (\ref{efolding}),
do not depend on $\beta$. Therefore,  since we can use  
\al{
\bar{V}_\text{inf}(S) &=
\bar{V}(S,\sigma_\text{v})=\frac{\beta^2}{M_\text{Pl}^4}\,V_\text{inf}(S,\sigma)
\label{VinfHat}
}
 to calculate these slow role parameters,
 the $\beta$ independence  
of $\bar{V}'_\text{inf}/\bar{V}_\text{inf}$ and  
$\bar{V}''_\text{inf}/\bar{V}_\text{inf}$
appearing in the slow role parameters 
becomes trivial.

So far the above discussion on the $\beta$ independence is exact.
There is in fact an origin of the $\beta$ dependence in
the slow role parameters: The function $F(S)$,
that is defined in Eq.~(\ref{field_norm}) and  is used to define
the canonically normalized field $\hat{S}$ 
in Eq.~(\ref{Shat}),
enters in these parameters.
To see it more explicitly, we square the both sides of Eq.~(\ref{Shat}) and obtain
\begin{align}
\frac{F(S)^2}{M^2_\text{Pl} } =\beta^{-1}\times
 \frac{ 1+Z_\sigma^{-1}\, 
 \sigma'^2_\text{v} }{ (1+4 A )\,\bar{B} } + \frac{3}{2} \frac{ \big(\, (1+4A )\,\bar{B}' 
+ 4A'\bar{B} \,\big)^2 }{ (1+4A )^2 \, \bar{B}^2 } \,,
\label{field_norm2}
\end{align}  
where $A(S,\sigma)$ is  independent of $\beta$, because
\al{
 A(S,\sigma) &= \frac{4G(S)\, 
 U(S,\sigma)}{ B(S)^2M_\text{Pl}^4 } 
 = \frac{4\bar{G}\,U(S,\sigma)}{\bar{B}(S)^2 } \,.
 }
Since the wave function renormalization $Z_\sigma^{-1}~(\simeq 0.3)$
and $\sigma_\text{v}$ are also independent of  $\beta$, 
the $\beta$ dependence of $F(S)$ can be simply factorized 
as we see in the first term of Eq.~(\ref{field_norm2}).
Note that  in our parameter space 
the ratio of the first term without $\beta$ to
 the second term is of order $10^{-1}$ and  
 $\beta\gsim 10^2$. Consequently, the $\beta$ dependence in $F(S)$ becomes 
negligibly small.
We therefore shall ignore the  first term in Eq.~(\ref{field_norm2}) 
in performing a parameter scan, so that  the independent parameters
are $\lambda_S,y$ and $\bar{\gamma}$.
 
 As announced, we impose the constraint (\ref{As_constraint})
 on $A_s$, which explicitly depends on $\beta$. Fortunately,
 this dependence is so simple, that it can be absorbed as
\al{
\bar{A}_s &=\beta^2 A_s
= \frac{\beta^2 V_{\rm inf}(S_*)}{24\pi^2\,\varepsilon_*\,M_{\rm Pl}^4} 
    =  
    \frac{ \bar{V}_{\rm inf} (S_*)}{24\pi^2\,\varepsilon_*}\,,
\label{rewriteAs}
}
where  $\bar{V}_{\rm inf}$ is defined in Eq.~(\ref{VinfHat}).
Obviously, $\bar{A}_s$ does not depend on  $\beta$.
 Thus, we calculate first $n_s, r, N_e$ and $\bar{A}_s$   for a given set of
  $\lambda_S,y,\bar{\gamma}$, and using the constraint on $A_s$
  (\ref{As_constraint}) we determine  the value of $\beta$ 
  from $\beta=10^{5}(\bar{A}_s/e^{3.044})^{1/2}$ and then
  $\gamma$ from   $\gamma=\bar{\gamma}\beta^2$.

 \subsubsection{Result}
 
\begin{figure}[ht]
	\begin{center}
		\includegraphics[width=5.63in]{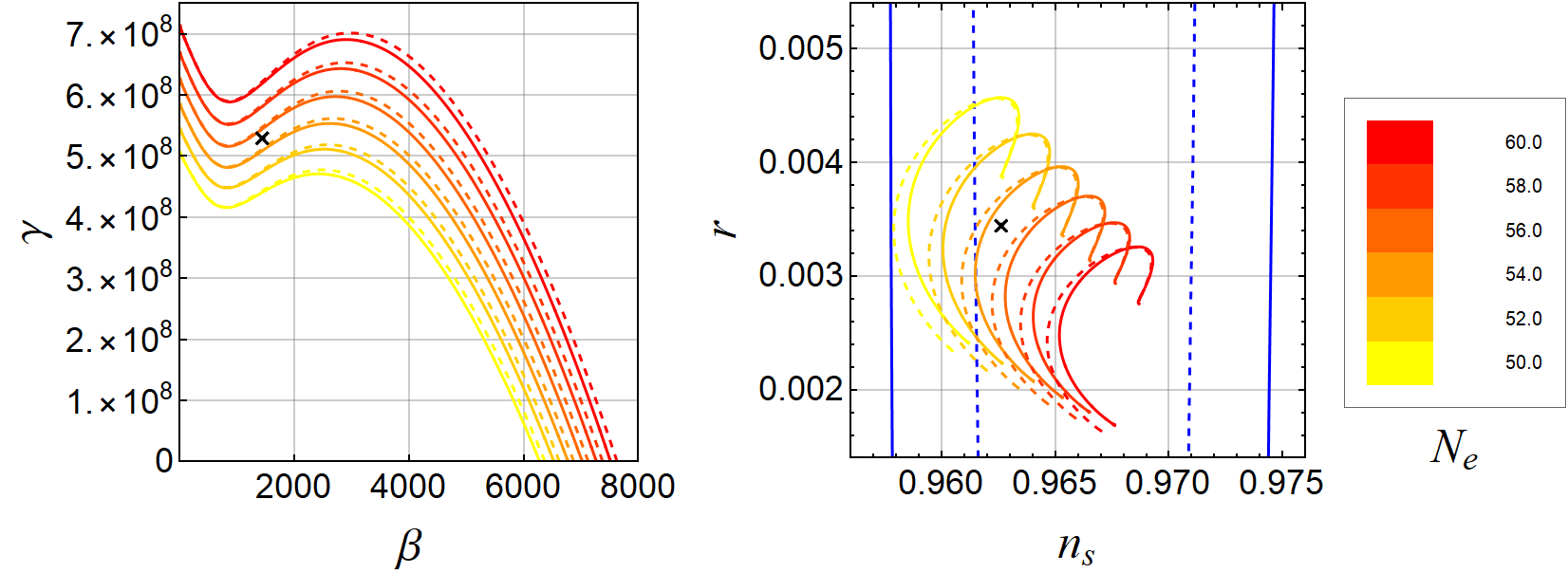}\hspace{2mm}
		\caption{The prediction on the inflationary parameters
		for $\lambda_S=1.20\times 10^{-2}$ 
		and $y=4.00\times 10^{-3}$ (solid line) and $4.00\times 10^{-4}$
		(dashed line), while the color represents $N_e$.
		The benchmark point (\ref{bench1-2}) is marked by a cross.
		Left: $\beta$ and $\gamma$ that satisfy
		Eq.~(\ref{As_constraint}).	
			Right: The prediction in the $n_s-r$ plane, where
		the solid (dashed) blue lines are two (one) $\sigma$ constraint
		by Planck \cite{Aghanim:2018eyx,Akrami:2018odb}.
}
		\label{lambdadot012}
	\end{center}
\end{figure}

\begin{figure}[ht]
	\begin{center}
		\includegraphics[width=5.63in]{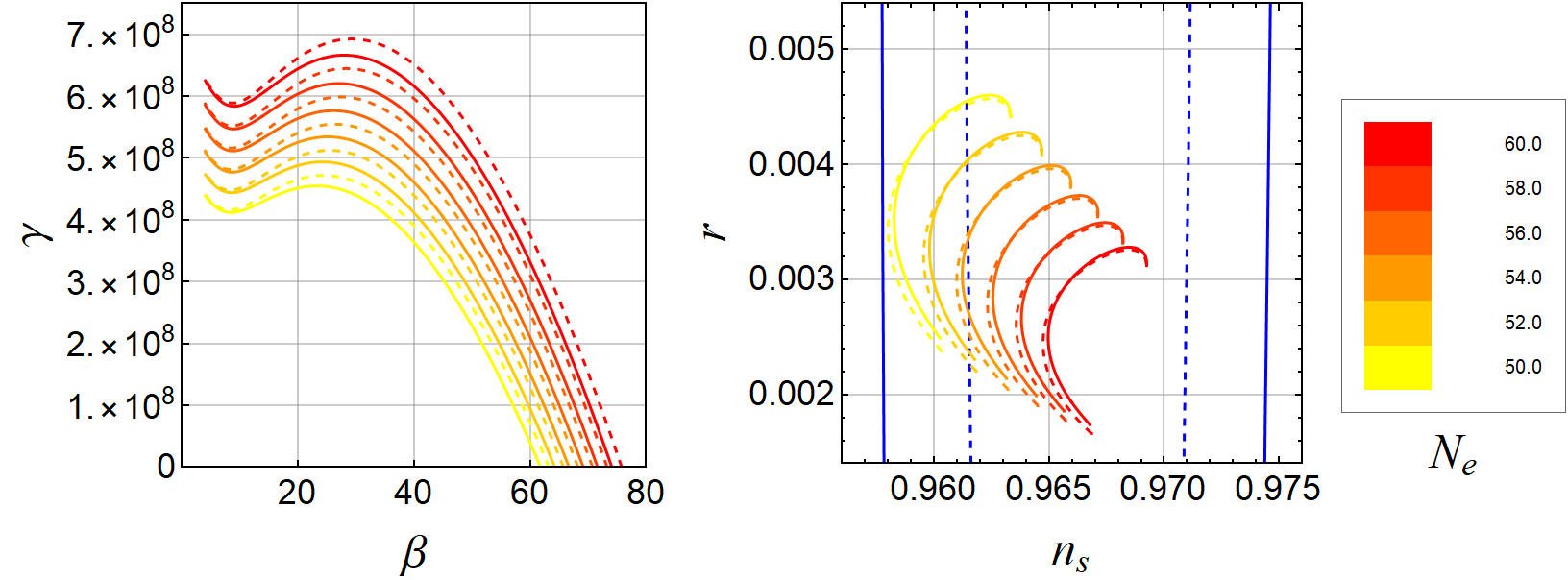}\hspace{2mm}
		\caption{The same as 
		for  Fig. \ref{lambdadot012} for $\lambda_S=1.20\times 10^{-6}$.
}
		\label{lambdadot012_6}
	\end{center}
\end{figure}

\begin{figure}[ht]
	\begin{center}
		\includegraphics[width=5.63in]{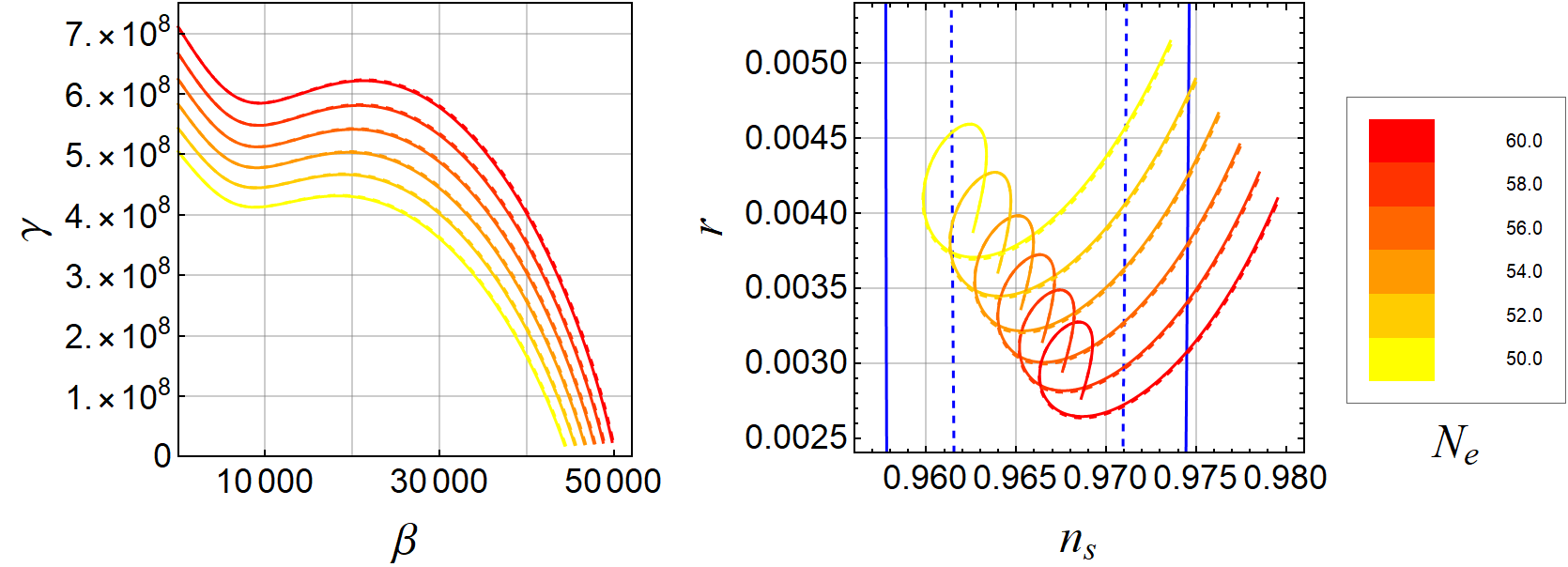}\hspace{2mm}
		\caption{The same as 
		for  Fig. \ref{lambdadot012} for $\lambda_S=1.20$.
}
		\label{lambdadot12}
	\end{center}
\end{figure}

The results are shown in Figs.~\ref{lambdadot012},
~\ref{lambdadot012_6},  and~\ref{lambdadot12} 
for  $\lambda_S=1.20\times 10^{-2},~1.20\times 10^{-6}$, and $1.20$, respectively.
In the left panels we show the values for $\beta$ and $\gamma$, while
in the right panels the corresponding values of $n_s$ and $r$
together with $N_e$  are presented.

If we vary $\bar{\gamma}=\gamma/\beta^2$ with $\ln(10^{10}A_s)=3.044$ fixed ,
we have a line in the  $\gamma-\beta$ plane as we can see in the left panels.
For a given set of $\lambda_S, y$ and $N_e$, the prediction
becomes a line in the $n_s-r$ plane, because we vary 
$\bar{\gamma}$.
Therefore, there is a one-to-one correspondence between 
the lines in the left panels and right panels.
The $y$ dependence of $n_s$ and $r$ is indeed small, but
$\Lambda_H/M_\text{Pl}$ is quite different as we can see from 
Eqs. (\ref{bench1-1}) and (\ref{bench2-1}).
By comparing Fig. \ref{lambdadot012}
with Fig. \ref{lambdadot12}, we can also see  that $n_s$ becomes larger 
if $\lambda_S$ becomes larger.
The solid (dashed) blue lines in the right panels indicates  the 
two-(one-) $\sigma$ constraint
		by Planck \cite{Aghanim:2018eyx,Akrami:2018odb}.
We see that the model predictions are in good agreement 
with the observed data.		
\section{Dark matter}\label{sec4}
\subsection{Mass spectrum and dark matter candidate}
\label{Mass spectrum}
\begin{figure}[ht]
\begin{center}
\includegraphics[width=3in]{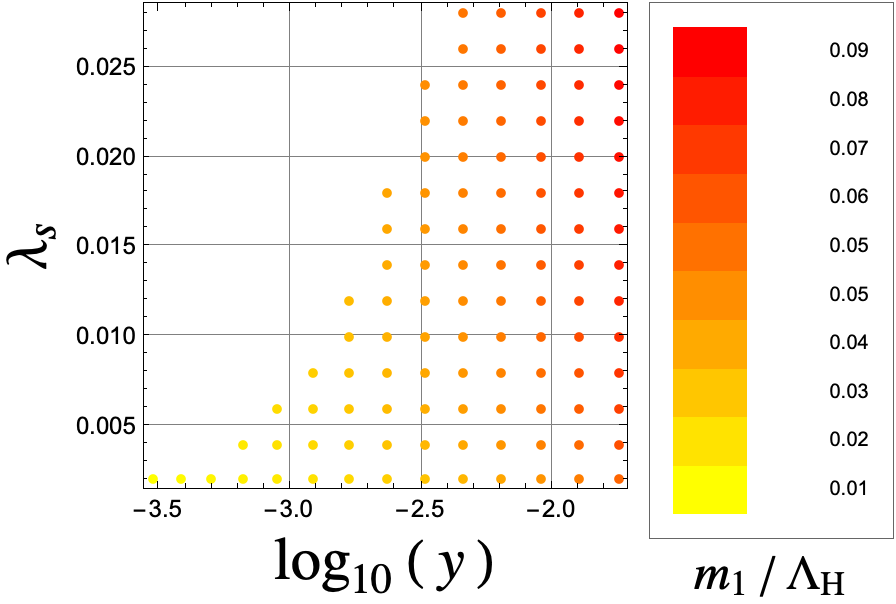}\hspace{0.5cm}
\includegraphics[width=3in]{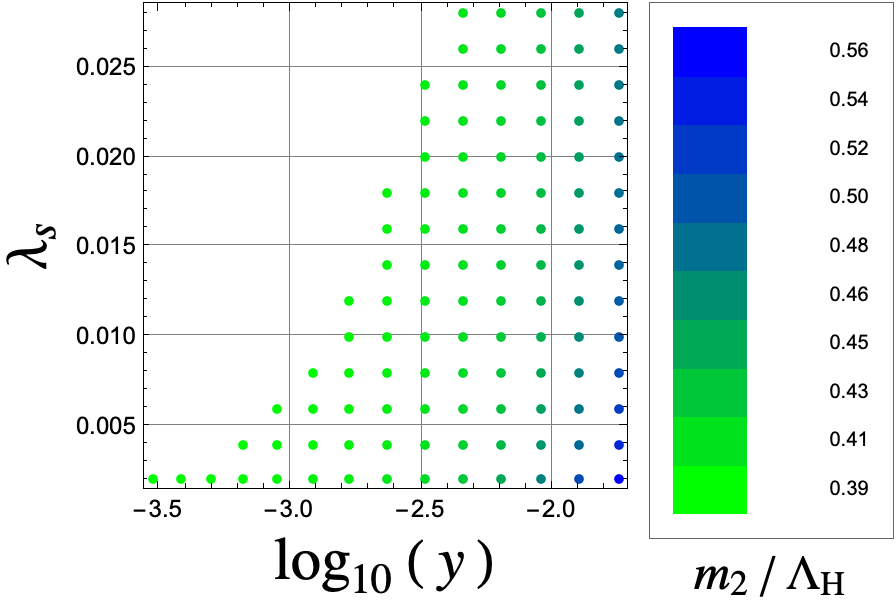}
\caption{The parameter space in the $y-\lambda_S$ plane
that we consider.
The data points are so chosen that  $ m_{\phi} > m_1$ 
is satisfied,
where $ m_{\phi}$ is calculated from Eq.~(\ref{GammaDM}).
 The size of 
$m_1/\Lambda_H$  and $m_2/\Lambda_H$  is shown with a color graduation.
In the most of the parameter space the mixing between $S $ and $\sigma$ 
is negligibly small, and $m_1\simeq \tilde{m}_S=\sqrt{3\lambda_3} v_S$
is satisfied.
}
\label{spectrum}
\end{center}
\end{figure}

Once the VEVs of $\sigma$ and $S$ are obtained, 
the scalar masses can be calculated by integrating out
 the hidden fermions.
These CP even scalars mix with each other, and 
the corresponding two point functions 
at the one-loop level $\Gamma_{AB} (A,B=S,\sigma)$
in the $SU(3)_V$ flavor symmetry limit are given by
\cite{Holthausen:2013ota,Kubo:2014ida,Ametani:2015jla}.
 \setcounter{equation}{0}
\renewcommand\theequation{4.\arabic{equation}}
\begin{align}
\Gamma_{SS}(p^2)&=p^2-3\lambda_{S}\left<S\right>^2
-y^2 3n_cI_{\varphi^2}(p^2,M,\Lambda_{\mathrm{H}})\,, \nn\\
\Gamma_{S\sigma}(p^2)&=-y\left(1-\frac{G_D\left<\sigma\right>}{4G^2}\right)3n_cI_{\varphi^2}(p^2,M,\Lambda_{\mathrm{H}})\,, 
\label{GammaSs}\\
\Gamma_{\sigma\sigma}(p^2)&=-\frac{3}{4G}+\frac{3G_D\left<\sigma\right>}{8G^3}-\left(1-\frac{G_D\left<\sigma\right>}{4G^2}\right)^23n_cI_{\varphi^2}(p^2,M,\Lambda_{\mathrm{H}}) 
+\frac{G_D}{G^2}3n_cI_V(M,\Lambda_{\mathrm{H}})\,,\nn
\end{align}
where we have neglected
 the Higgs portal coupling, and the loop functions are defined as
\begin{align}
 \label{propagator for sigma}
  I_{\varphi^2}(p^2,M,\Lambda) &= \int_{\Lambda} \frac{d^4 k}{i(2\pi)^4}\frac{\mathrm{Tr}(\Slash{k}+\Slash{p}+M)(\Slash{k}+M)}{((k+p)^2-M^2)(k^2-M^2)}\,, \nn\\
  I_{V}(M,\Lambda) & = \int_{\Lambda}  \frac{d^4 k}{i(2\pi)^4}\frac{M}{(k^2-M^2)}=-\frac{1}{16\pi^2}M\left[ \Lambda^2-M^2\ln \left( 1+\frac{\Lambda^2}{M^2}\right) \right]\,.
\end{align}
The mixed fields $(S, \sigma)$ 
and the diagonalized fields  $(S_1,S_2)$
corresponding to the mass eigenstates are related by 
\al{
\left(\begin{array}{c}S\\ \sigma
\end{array}\right) &=
\left(\begin{array}{cc}
\xi_S^{(1)} & \xi_S^{(2)} \\ 
  \xi_\sigma^{(1)} & \xi_\sigma^{(2)} 
\end{array}  \right)~\left(\begin{array}{c}S_1\\S_2 
\end{array}\right)\,,
\label{mixingM}
}
where we denote the mass of $S_i$ by $m_i$.
The mixing parameters $\xi_{S,\sigma}^{(i)}$
and $m_i$ can  be obtained by solving
\al{
\sum_{B=S,\sigma}\Gamma _{AB}\,(m^2_{i})\,\xi_{B}^{(i)}&=0\,. 
}
For the benchmark point given in Eq.~(\ref{bench1}), we find
~\footnote{The mixing matrix in Eq.~(\ref{mixingM})
is not an orthogonal matrix. This means that after diagonalization 
one  has to perform an appropriate wave function renormalization
to define canonically normalized fields. 
Since the mixing is very small and we are mostly
interested in $S$ being inflaton, we ignore this procedure here.}:
\al{
m_1/\Lambda_H &\simeq0.044\,,\quad m_2/\Lambda_H \simeq 0.41\,,\quad \nn\\
  \xi_S^{(1)}&\simeq 1.00\,,\quad   \xi_\sigma^{(1)}\simeq 0.01\,,\quad 
    \xi_S^{(2)}\simeq -0.006\,,\quad   \xi_\sigma^{(2)}\simeq 1.00\,.
    \label{m12}
}
Therefore, the mixing is very small, and we find that $m_1$ can be well approximated
by $\tilde{m}_S=\sqrt{3\lambda_S} v_S$.
We have calculated $m_1/\Lambda_H$ and 
$m_2/\Lambda_H$ as a function of $y$ and $\lambda_S$ 
for an area in the parameter space which is relevant for our purpose.
This is plotted in Fig.\ref{spectrum}, where the size of 
$m_i/\Lambda_H$ is shown with a color graduation.
(The data points  satisfy $ m_{\phi} > m_1$, 
where $ m_{\phi}$ is calculated from Eq.~(\ref{GammaDM}).)
We find that the mixing between $S$ and $\sigma$ in the parameter space
is very small as it is the case for the example shown in Eq.~(\ref{m12}).
We also find that in the most of the area of  the parameter  space 
$m_1 \ll  m_2$ and $m_1\simeq \sqrt{3\lambda_S} v_S$
are satisfied. Therefore, $\sigma$ will not play any role for
our discussion below, and we use the approximate formula
$m_1\simeq \tilde{m}_S$ in the following discussions.

Due to the vector-like flavor symmetry
(i.e. $SU(3)_V$ or its subgroup), the dark meson,
the CP-odd scalar $\phi_a$  in Eq.~(\ref{varphi}),
is a good DM candidate.
 The two point function at the one-loop level for  $\phi_a$ is written as
\cite{Holthausen:2013ota,Kubo:2014ida,Ametani:2015jla}
\begin{align}
\label{GammaDM}
\Gamma_{\phi}(p^2)&=-\frac{1}{2G}+\frac{G_D\left<\sigma\right>}{8G^3}+\left(1-\frac{G_D\left<\sigma\right>} {8G^2}\right)^22n_cI_{\phi^2}(p^2,M,\Lambda_{\mathrm{H}}) +\frac{G_D}{G^2}n_cI_V(M,\Lambda_{\mathrm{H}})\,,
\end{align}
where the loop function $ I_{\phi^2}(p^2,M,\Lambda)$ is given by
\begin{align}
  I_{\phi^2}(p^2,M,\Lambda) &= \int_{\Lambda}  \frac{d^4 k}{i(2\pi)^4}\frac{\mathrm{Tr}(\Slash{k}-\Slash{p}+M)\gamma_5(\Slash{k}+M)\gamma_5}{((k-p)^2-M^2)(k^2-M^2)}\,,
\end{align}
and  $m_{\phi}$ is defined by
$\Gamma_{\phi}(m_{\phi}^2)=0$.
For the benchmark point we find 
$m_\phi/\Lambda_H \simeq 0.06 > \tilde{m}_S/\Lambda_H$.
This implies that, if $S$ is inflaton, it cannot decay into
the dark meson~\footnote{The dark meson $\phi_a$ can also be produced by the annihilation of $N$. However, as shown in the next subsection, its cross section is very small due to the small $y_M$. Therefore we consider the DM production from inflaton decay.
}. As we see from Fig. (\ref{spectrum}),
$m_\phi > \tilde{m}_S$ is satisfied in the most of the parameter space,
especially for large $y (\gsim 0.003)$.
We therefore break $SU(3)_V$ down to $SU(2)_V\times U(1)$
 and  assume a
 hierarchy in the Yukawa couplings   \cite{Ametani:2015jla}: 
 \al{
 {\bm y} &= \mbox{diag.}\,(y_1\,,\,y_1\,,\,y_3\,)~
 \mbox{with}~ y_1=y_2< y_3,
 }
 where  ${\bm y}$ is the Yukawa matrix in
 the hidden sector described by the Lagrangian (\ref{LH}).
Under this assumption,
the dark mesons fall into three categories, $\tilde{\pi}=\left\{ \tilde\pi^\pm,~\tilde{ \pi}^0 \right\}, 
\tilde{K} =\left\{\tilde{K}^\pm,~\tilde{K}^0, ~\bar{\tilde{K}}^0 \right\}$
and $\tilde{\eta}$.
Here the dark mesons are named like the real-world mesons:
 \begin{align}
\tilde{\pi}^\pm &\equiv (\phi_1\mp i\phi_2)/\sqrt{2}~,~~
\tilde{ \pi}^0 \equiv \phi_3~, ~\nn\\
 \tilde{K}^\pm & \equiv (\phi_4\mp i  \phi_5)/\sqrt{2}~,
 ~~\tilde{K}^0(\bar{\tilde{K}}^0) \equiv
  (\phi_6+(-) i \phi_7)/\sqrt{2}~,~~ \tilde{\eta}^8 \equiv \phi_8\,,
  \label{mesons}
  \end{align}
where $\tilde{\eta}^8$ will mix with $\tilde{\eta}^0=\phi_0
$ to form the mass eigenstates
$\tilde{\eta}$ and $\tilde{\eta}'$. 
The states in the same category have the same mass, 
$m_{\tilde{\pi}^0}=m_{\tilde{\pi}^\pm} ( \equiv  m_{\tilde{\pi}}) $ and 
$m_{\tilde{K}^\pm}=m_{\tilde{K}^0}=m_{\bar{\tilde{K}}^0} ( \equiv m_{\tilde{K}} )$,
with $m_{\tilde{\pi}} < m_{\tilde{K}} < m_{\tilde{\eta}}$.
As we will see in the next section, the dark meson mass has to be several orders of magnitude smaller than $\tilde{m}_S$, such that we can obtain a realistic DM abundance.
The dark meson mass decreases as the Yukawa coupling  decreases.
However, if we decrease  by several orders of magnitude, 
the cutoff $\Lambda_H$ increases accordingly
and  may exceed $M_\text{Pl}$ by various  orders of magnitude, 
which we want to  avoid.
A nice way out exits if there is a parameter space in which 
$m_{\tilde{\pi}} \ll \tilde{m}_S < m_{\tilde{K}} < m_{\tilde{\eta}}$ is realized,
as we will argue in the next section when discussing DM relic abundance.

\begin{figure}[ht]
\begin{center}
\includegraphics[width=3.5in]{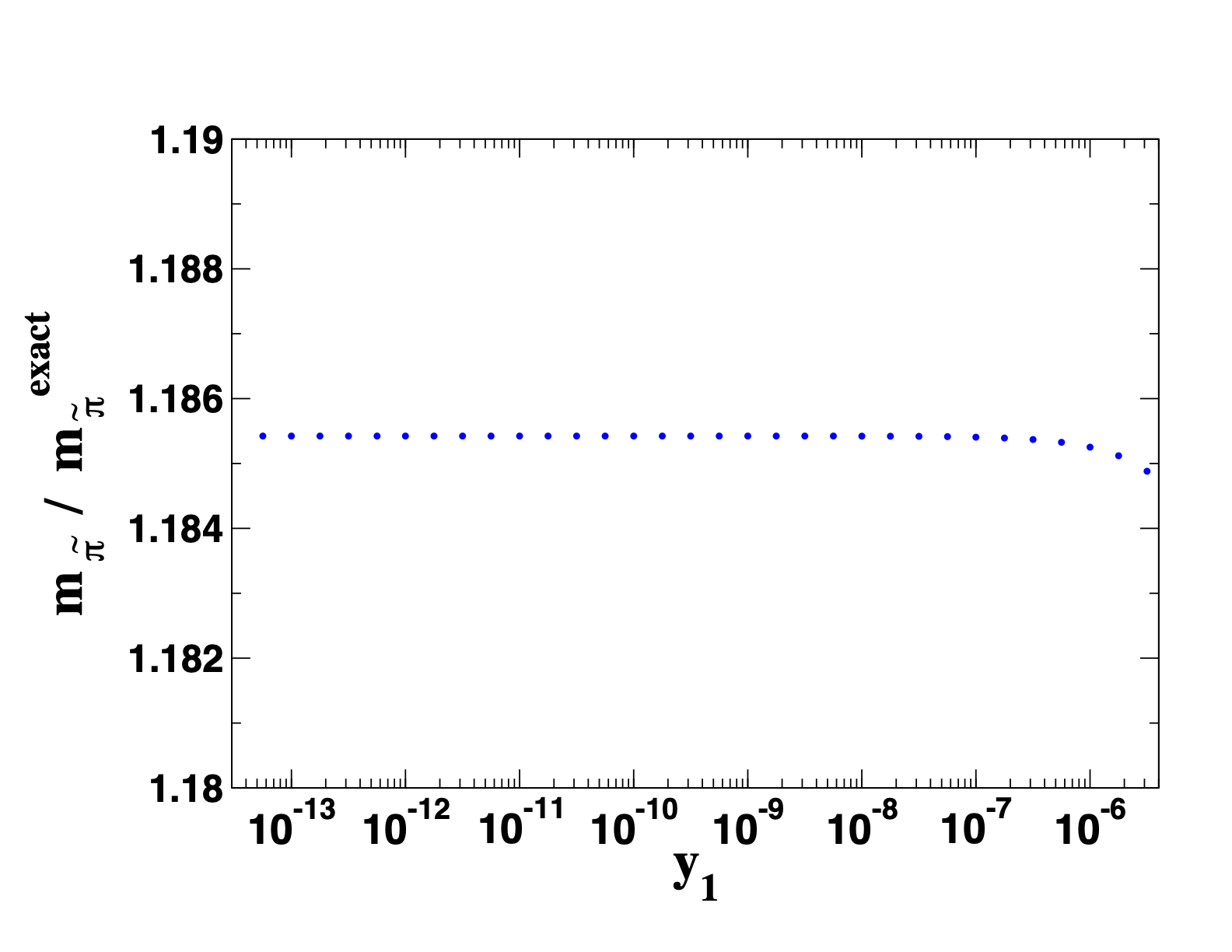}
\caption{The ratio $m_{\tilde{\pi}}/m_{\tilde{\pi}}^\text{  exact}$ 
against $y_1$,
where $m_{\tilde{\pi}}$ is calculated by using Eq.~(\ref{mpi1}),
and $m_{\tilde{\pi}}^\text{  exact}$
 is calculated by using the formula  of Ref. \cite{Ametani:2015jla}
for the $SU(2)_V\times U(1)$ case.}
\label{y1-ratio}
\end{center}
\end{figure}

If  $SU(3)_V$ 
is broken to the $SU(2)_V\times U(1)$, Eq.~(\ref{GammaDM}) is no longer
applicable to obtain the dark meson mass.
Fortunately, there exists a good approximation \cite{PhysRev.175.2195} 
\al{
m_{\tilde{\pi}}^2/m_{\phi}^2 &\simeq m_u/m_q\simeq y_1/y\,,
\label{mpi1}
}
where $m_u$ is the current quark mass in the $SU(2)_V\times U(1)$ case, and
$m_q$, $m_{\phi}$ and $y$ are the current quark mass,
the dark meson mass and the Yukawa coupling, respectively,
in the $SU(3)_V$ limit.
 In Fig. \ref{y1-ratio} we show $m_{\tilde{\pi}}/m_{\tilde{\pi}}^\text{  exact}$
for  $3\times 10^{-14} < y_1<  3\times 10^{-6}$ (while $\tilde{m}_S < m_{\tilde{K}}$ is satisfied),
where $m_{\tilde{\pi}}^\text{  exact}$
 is the dark pion mass calculated by using the formula 
of Ref. \cite{Ametani:2015jla}
for the $SU(2)_V\times U(1)$ case.
As we see from Fig. \ref{y1-ratio} 
the difference between $m_{\tilde{\pi}}$ and 
$m_{\tilde{\pi}}^\text{  exact}$ is less than $20$\% for a wide range of $y_1$,
and we shall use this approximation.

The message of this section is that there exits a sufficiently large parameter
space, in which
$m_{\tilde{\pi}} \ll \tilde{m}_S < m_{\tilde{K}} < m_{\tilde{\eta}}$ can be realized.

\begin{figure}[ht]
\begin{center}
\includegraphics[width=3in]{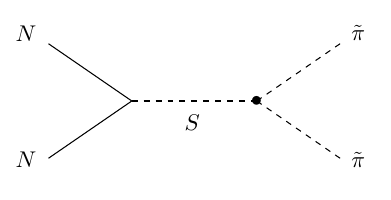}
\caption{The annihilation process $N\, N\leftrightarrow \tilde{\pi}\,\tilde{\pi} $,
where the $N-N-S$ coupling is $y_M=m_N/v_S$ and 
the $\tilde{\pi}-\tilde{\pi}-S$ effective coupling
(indicated by a bullet)  is denoted $G_{\tilde{\pi}\tilde{\pi} S}$ in the text,
which can be calculated from the diagrams shown in Fig. \ref{pipiS}
\cite{Ametani:2015jla}.
}
\label{NNpipi}
\end{center}
\end{figure}

\begin{figure}[ht]
\begin{center}
\includegraphics[width=2.3in]{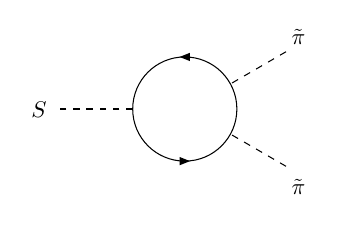}\hspace{0.5cm}
\includegraphics[width=2.3in]{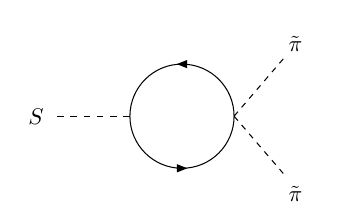}
\caption{One-loop diagrams contributing to the effective coupling
$G_{\tilde{\pi}\tilde{\pi} S}$, which has been calculated in
Ref. \cite{Ametani:2015jla}.
}
\label{pipiS}
\end{center}
\end{figure}

\subsection{Dark matter relic abundance}
Dark mater can be produced during or after the reheating phase, see e.g.~\cite{Chung:1998rq,Allahverdi:2002nb,Garcia:2020eof}.
In the following discussions we assume that the Yukawa couplings
$y_1$ and $y_3$ are so chosen that the mass hierarchy 
$2m_{\tilde{\pi}} < \tilde{m}_S < m_{\tilde{K}}<m_{\tilde{\eta}}$ is satisfied.
Under this assumption the inflaton $S$
can decay only into a pair of $\tilde{\pi}$ - which is our DM - but not into the other mesons,
and therefore they are
 not produced during the reheating stage and later~\cite{Allahverdi:2002nb}.
$\tilde{\pi}$  can also be produced 
by the annihilation process $N\, N\leftrightarrow \tilde{\pi}\,\tilde{\pi} $.
The corresponding $S$ channel diagram is shown in 
Fig.~\ref{NNpipi}. As we see from this diagram, the annihilation
cross section is
proportional to $y_M^2\,G_{\tilde{\pi}\tilde{\pi} S}^2$, where
$G_{\tilde{\pi}\tilde{\pi} S}^2$ is 
the effective $S-\tilde{\pi}-\tilde{\pi}$ coupling.
Since $y_M^2\sim m_N^2/v_S^2\sim 10^{-18}$, the $\tilde{\pi}$
production through 
the annihilation process is negligibly small
compared with that trough the decay which is proportional only 
to $G_{\tilde{\pi}\tilde{\pi} S}^2$.
Therefore, we ignore this annihilation process 
and take into account only  the decay of $S$ into  $\tilde{\pi}$, with
the decay width
\al{
\gamma_{\tilde{\pi}} &=
\frac{3 \,G_{\tilde{\pi}\tilde{\pi} S}^2}{16\pi \tilde{m}_S} 
\sqrt{1-\frac{4 m_{\tilde{\pi}}^2} {\tilde{m}_S^2}  }\,,
\label{gammapi}
}
where the effective coupling  is calculated  in Ref. \cite{Ametani:2015jla}
from the diagram shown in Fig. (\ref{pipiS}) and is found to be
$G_{\tilde{\pi}\tilde{\pi}S}/\Lambda_H
\simeq -0.012\,y_1$ for $y_1\lsim 5\times 10^{-4}$.

In this way we arrive at a system, which consists of 
only the inflaton $S$ and the $\tilde{\pi}$. 
The evolution of the number densities,  $n_S$ and $n_{\tilde{\pi}}$, can be described
by the coupled Boltzmann equations~\cite{Chung:1998rq}
\al{
\frac{d n_S}{dt} &= -3 H n_S-\Gamma_S\, n_S \,,
\label{nS}\\
\frac{d n_{\tilde{\pi} }}{dt}&= -3 H n_{\tilde{\pi}}
+\gamma_{\tilde{\pi}} \,n_S \,,
\label{npi}
}
where 
$\Gamma_S$ is the total decay width of $S$.
Eq.~(\ref{nS})  can be simply solved \cite{Kolb:1990vq}:
\al{
n_S (a) &= \frac{\rho_\mathrm{end}}{\tilde{m}_S}\,\left[
\frac{a_\mathrm{end}}{a}\right]^3\, e^{-\Gamma_S\,(t-t_\mathrm{end})}\,,
\label{nSa}
}
where $a$ is the scale factor at $t > t_\mathrm{end}$, $a_\mathrm{end}$
is $a$ at the end of inflation $t_\mathrm{end}$ and 
$\rho_\mathrm{end}=\rho_S(a_\mathrm{end})=\tilde{m}_S n_S(a_\mathrm{end})$
is the inflaton energy density at $t_\mathrm{end}$.
Then we insert the solution 
(\ref{nSa}) into Eq.~(\ref{npi})  to find
\al{
n_{\tilde{\pi}} (a) &=
B_{\tilde{\pi}} \,\frac{\rho_\mathrm{end}}{\tilde{m}_S}\,
\left[\frac{a_\mathrm{end}}{a}\right]^3\,
\left(1- e^{-\Gamma_S\,(t-t_\mathrm{end})}\right)~\mbox{with}~
B_{\tilde{\pi}} =\frac{\gamma_{\tilde{\pi}}}{\Gamma_S}\,,
}
from which  we obtain
 the DM relic abundance
\al{
\Omega_{\tilde{\pi}} h^2 &=
m_{\tilde{\pi}}\,B_{\tilde{\pi}} \,\frac{\rho_\mathrm{end}}{\tilde{m}_S}\,
\left[\frac{a_\mathrm{end}}{a_0}\right]^3\, 
\frac{1}{3  M_\mathrm{Pl}^2 (H_0/h)^2}\,,
}
where  $a_0=1$  and $H_0= h\,2.1332\times 10^{-42}$ GeV
with $h\simeq 0.674$ \cite{Aghanim:2018eyx}
stand for the present value of the scale factor and the
Hubble parameter, respectively.
To proceed, we write the ratio $a_\mathrm{end}/a_0$ as
\al{
\frac{a_\mathrm{end}}{a_0} &=R_\mathrm{rad}~F_\text{LA}~\left(\frac{\sqrt{3} \,H_0}{\rho_\mathrm{end}^{1/4}}  \right)\,,
\label{aend}
}
where
\al{
R_\mathrm{rad}&=\left(\frac{a_\mathrm{end}}{a_\mathrm{RH}}\right) ~
\left(\frac{\rho_\mathrm{end}^{1/4}}{\rho_\mathrm{RH}^{1/4}}\right) 
= \left(\frac{\rho_\mathrm{RH}}{\rho_\mathrm{end}}\right)^{
\frac{1-3\bar{w}}{12(1+\bar{w})}}
\label{Rrad}
}
with  $\bar{w}$ being the average equation of state~\cite{Martin:2010kz}, and 
\al{
F_\text{LA} &=\frac{a_\mathrm{RH}\,\rho_\mathrm{RH}^{1/4}}{\sqrt{3} a_0 \,H_0}
=\exp\left( 66.89- \frac{1}{12}\ln g_\mathrm{RH}\right)\,,\label{FLA}
}
with $g_\mathrm{RH}$ being the relativistic degrees of freedom at the end of reheating~\cite{Lozanov:2017hjm,Akrami:2018odb}.
The average equation of state $\bar{w}$ 
is zero in the present case, because  $V_\text{inf}(S)$ in Eq.~(\ref{Vinf})
behaves as $V_\text{inf}(\hat{S})\sim \hat{S}^2$ near the potential minimum,
where $\hat{S}$ is the canonically normalized field and defined in Eq.~(\ref{Shat}).
Consequently, the $\rho_\mathrm{end}$ dependence in $\Omega_{\tilde{\pi}}
h^2$ cancels.
Further, introducing  the reheating temperature $T_\mathrm{RH}$ as
\al{
\rho_\mathrm{RH}
&= \frac{\pi^2}{30}\,g_\mathrm{RH} \,T_\mathrm{RH}^4\,,
\label{TRH}
}
we  find~\cite{Allahverdi:2002nb}. 
\al{
\Omega_{\tilde{\pi}} h^2 &=
\sqrt{3} \exp(3\times 66.89)\,\frac{B_{\tilde{\pi}} \,H_0 h^2}{M_\mathrm{Pl}^2}\,
\left(\frac{\pi^2}{30}\right)^{1/4}\,
\left(\frac{m_{\tilde{\pi}}}{\tilde{m}_S}\right)\,T_\mathrm{RH}\,\nn\\
&\simeq 2.04\times 10^{8}\,B_{\tilde{\pi}} \left(\frac{m_{\tilde{\pi}}}{\tilde{m}_S}\right) \,\frac{T_\mathrm{RH}}{1\,\mbox{GeV}}\,.\label{OmegaDM}
}
The branching ratio $B_{\tilde{\pi}}=\gamma_{\tilde{\pi}}/\Gamma_S$ can be obtained
from $\gamma_{\tilde{\pi}}$ 
given in Eq.~(\ref{gammapi})  together with  the assumption that
$1/\Gamma_S$ is the time scale at the end of the reheating phase~\cite{Kolb:1990vq,Chung:1998rq}, which means\ $1/H(a_\mathrm{RH})
=\left(\,3\, M_\mathrm{Pl}^2/\rho_\mathrm{RH}
\,\right)^{1/2}$. 

\subsection{Reheating temperature and DM relic abundance}
As we see from  Eq.~(\ref{OmegaDM}), we need to know the reheating temperature $T_\text{RH}$
to obtain an actual value of the DM relic abundance $\Omega_{\tilde{\pi}} h^2$.
Fortunately, it is possible ~\cite{Liddle:2003as,Martin:2010kz} to constrain
the reheating phase and hence
 $T_\text{RH}$ for a given inflation model without specifying  reheating mechanism. We will follow this idea to find consistent values for $T_\text{RH}$
 for our model.

 The basic unknown quantities during the reheating phase are
the expansion rate 
$a_\mathrm{end}/a_\mathrm{RH}$ 
and the energy density $\rho_\mathrm{RH}$
at the end of reheating.
These uncertainties can be taken into account 
in $R_\mathrm{rad}$~\cite{Martin:2010kz},
 which has been already introduced  in Eq.~(\ref{Rrad}).
We then consider the ratio $a_\mathrm{end}/a_*$,
where $a_*=k_*/H_*$ is the scale factor at the time of CMB horizon exit,
$k_*$ is the pivot scale set by the Planck
mission \cite{Aghanim:2018eyx,Akrami:2018odb}, 
and $H_*$ is the Hubble parameter at $a=a_*$:
\al{a_\mathrm{end}/a_*=R_\mathrm{rad} \,
F_\text{LA} \left(\sqrt{3} H_*/\rho_\mathrm{end}^{1/4}\right)
\left(a_0 H_0/k_*\right)\,.
\label{a-ratio}
}
($F_\text{LA}  $ is given in Eq.~(\ref{FLA})).
Using Eq.~(\ref{a-ratio}) we find that the number of e-folds
$ N_e $ can be written as ~\cite{Martin:2010kz,Martin:2013tda}
\begin{equation}
\begin{split}
N_e &=\ln \left(\frac{a_\mathrm{end}}{a_*}\right)=
66.89-\frac{1}{12}\ln g_\mathrm{RH}+\frac{1}{12}\ln\left(\frac{\rho_\text{RH}}{\rho_\text{end}}\right)+\frac{1}{4} \ln \left(\frac{V_\mathrm{inf\,*}^2}{M_\mathrm{Pl}^4 \,\rho_\mathrm{end}}  \right) -\ln \left(\frac{k_*}{a_0 H_0}  \right)\,\\&=
66.80
-\ln \left(\frac{k_*}{a_0 H_0}  \right)+\frac{1}{4}
\ln \left(\frac{V_\mathrm{inf\,*}^2}{M_\mathrm{Pl}^4 \,\rho_\mathrm{end}}  
\right)-
\frac{1}{12}
\ln \left(
\frac{V_\mathrm{end}(3-\varepsilon_*)}{(3-\varepsilon_\mathrm{end})M_\text{Pl}^4}
\right)+\frac{1}{3}\ln \left(\frac{T_\text{RH}}{M_\text{Pl}}\right)\,,
\label{Ne}
\end{split}
\end{equation}
where  we have used
\al{\sqrt{3}H_* &=\frac{V_\mathrm{inf\,*}^{1/2}}{M_\text{Pl}}~~\mbox{and}~~
\rho_\mathrm{end} =
\frac{V_\mathrm{end}(3-\varepsilon_*)}{(3-\varepsilon_\mathrm{end})}\,,
}
and $V_\mathrm{end}=V_\mathrm{inf}(S_\mathrm{end})\,,
V_\mathrm{inf\,*}=V_\mathrm{inf}(S_*)$,
$\varepsilon_\mathrm{end}=\varepsilon(S_\mathrm{end})$, and
$\varepsilon_*=\varepsilon(S_*)$.
Note that the $g_\mathrm{RH}$ dependence in Eq.~(\ref{Ne}) disappears. 

\begin{figure}[t]
	\begin{center}
			\includegraphics[width=3.1in]{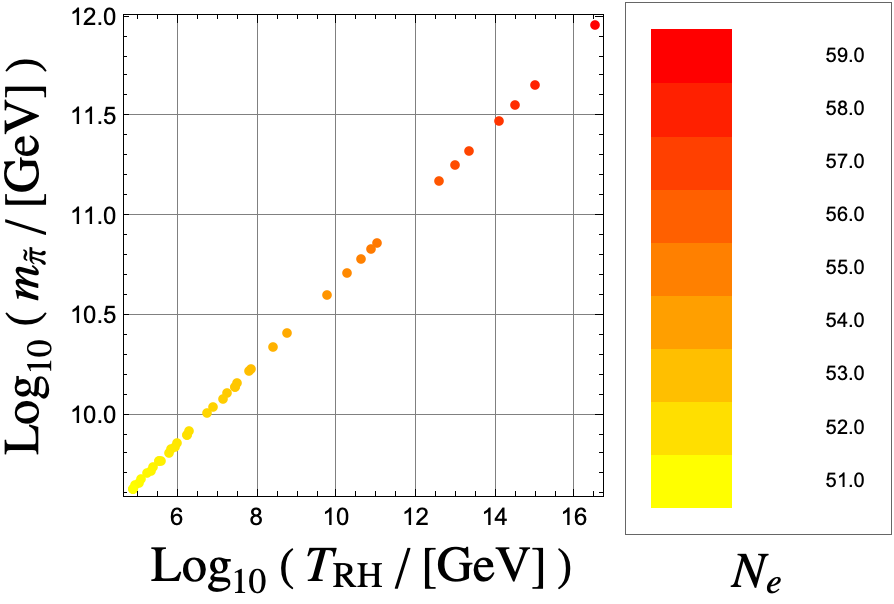}
						\hspace{2mm}
			\includegraphics[width=3.2in]{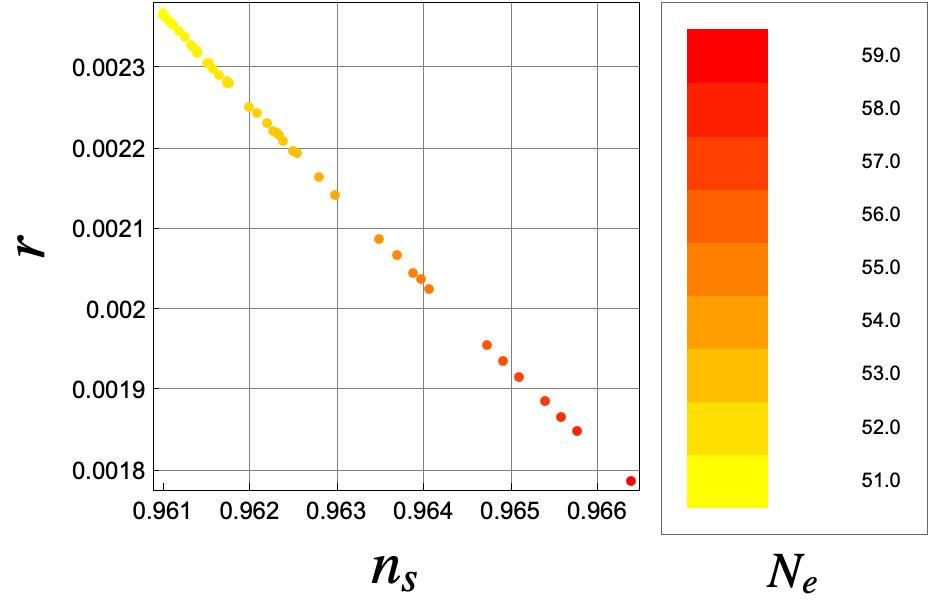}

		\vspace{-2mm}
		\caption{\footnotesize{Left: Dark matter mass $m_{\tilde{\pi}}$ against the 
		reheating temperature $T_\mathrm{RH}$.
		The colored points satisfy  
		 $\Omega_\pi h^2=0.1198\pm 0.0024\,(2\sigma)$,
		 where the color represents the e-foldings $N_e$.
	We have varied  $y_1$ and  $\gamma$
	for  fixed $y_3,\,\beta$ and $\lambda_S$
	at $4.00\times 10^{-3},\, 6.31\times 10^3$ and $1.2\times 10^{-2}$, respectively,
	such that the constraint on $A_s$ given in Eq.~(\ref{As_constraint}) is satisfied.
	  Note that the lower bound on $T_\mathrm{RH}$ 
	for a viable thermal leptogenesis with $m_N \gtrsim 2\times10^7$ GeV is about $10^9$ GeV~\cite{Giudice:2003jh}.
	Right: The same points as in the left panel are displayed in the $n_s-r$ plane.
			}
		}
		\label{TRH-DM1}
	\end{center}
\end{figure}

\begin{figure}[t]
	\begin{center}
						\includegraphics[width=3.1in]{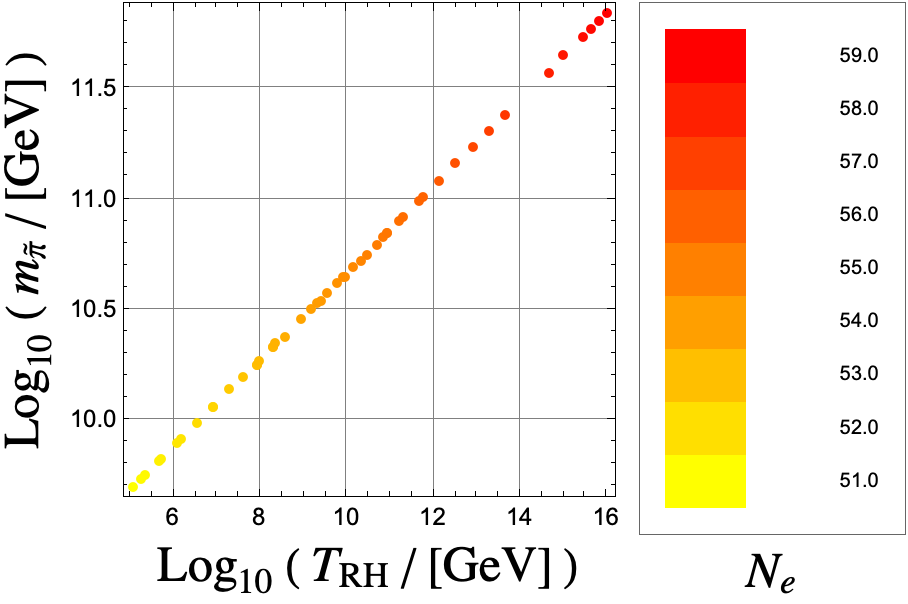}
						\hspace{2mm}
			\includegraphics[width=3.2in]{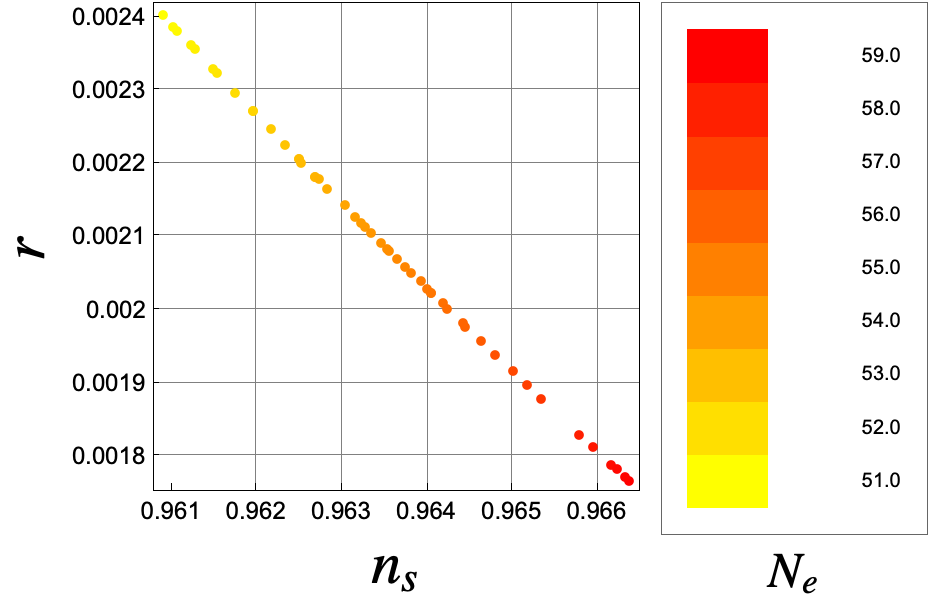}

		\vspace{-2mm}
		\caption{\footnotesize{The same as Fig.\ref{TRH-DM1}
		with $y_3=4.00\times 10^{-3},\, \beta=6.31\times 10^3, \,\gamma=1.26\times 10^8$,
		while $\lambda_S$ and $y_1$ are varied.}
		}
		\label{TRH-DM2}
	\end{center}
\end{figure}

For the benchmark point (\ref{bench1}) (see also 
(\ref{bench1-1}) and  (\ref{bench1-2})) with $y_3=4.00\times 10^{-3}$ and $y_1=4.47\times 10^{-13}$,
we find
 \al{
 \Lambda_H &\simeq 1.30\times 10^{17}\,\mbox{GeV}\,,\quad
 m_{\tilde{\pi}} \simeq 8.31\times 10^{10}\,\mbox{GeV}\,,\quad \tilde{m}_S\simeq 5.67\times 10^{15}\,\mbox{GeV}\,,
\nn\\
G_{\tilde{\pi}\tilde{\pi}S} &\simeq
-2.35\,y_1\times 10^{15}\,\mbox{GeV}
\simeq -1.05\times 10^{3}\,\mbox{GeV}\,,\\
T_\mathrm{RH} &\simeq  2.07\times 10^{11}\,\mbox{GeV}\,,\quad
B_{\tilde{\pi}} \simeq 1.90\times 10^{-16}\,,\nn
}
which gives
$\Omega_{\tilde{\pi}} h^2\simeq 0.119$.
In Figs.~\ref{TRH-DM1} and \ref{TRH-DM2} (left) we show the points in 
the $T_\mathrm{RH}-m_{\tilde{\pi}}$ plane,  
for which $\Omega_{\tilde{\pi}} h^2=0.1198\pm 0.0024\,(2\sigma)$
is obtained. 
Note that $m_{\tilde{\pi}}$ (\ref{mpi1}) and $\gamma_{\tilde{\pi}}$ (\ref{gammapi})
appearing in $\Omega_{\tilde{\pi}}h^2$ (\ref{OmegaDM}) strongly
depend on $y_1$, so that they are closely correlated.
We have varied   $y_1$ and  $\gamma\, (\lambda_S)$  for Fig.~\ref{TRH-DM1} 
(\ref{TRH-DM2})
	for  fixed $y_3,\,\beta$ and $\lambda_S\, (\gamma)$,
	such that the constraint (\ref{As_constraint}) is satisfied.
In the right panel we show the corresponding values of $n_s$ and $r$.

\section{Conclusion}\label{sec5}
We have followed
John Wheeler’s requirement \cite{Wilczek:1999be} that the fundamental equations 
should not contain any dimensionful parameter.
Accordingly,  we have started with a theory, which contains no dimensionful parameter at the classical level. Since there exits energy scales in the real word, 
they have to be generated.
We have two known mechanisms of “scalegenesis” at hand;
the Coleman-Weinberg mechanism   and 
the dynamical symmetry breaking by strong dynamics 
in nonabelian gauge theories.
The former mechanism
is based on improved perturbation theory, while the later one
uses nonperturbative effect in nonabelian gauge theories,
e.g., chiral symmetry breaking in QCD, 
which produces about 99 \%
 of the proton mass.
 
 In this paper we have assumed that the origin of the dimensionful
 parameters, i.e. the Planck mass and the electroweak scale
 including the right-handed neutrino mass, is 
  chiral symmetry breaking in a QCD-like theory, which
couples with  the visible sectors only via a real scalar
$S$, the mediator.
It is not only a mediator, but also inflaton,
which makes a Higgs-inflation-like scenario possible.
In fact the prediction of the CMB observables,
the scalar spectral index $n_s$ and the tensor-to-scalar ratio $r$,
is similar to that of the Higgs-inflation \cite{Bezrukov_2008}
 (or $R^2$ inflation \cite{Starobinsky:1980te,1981JETPL..33..532M,Starobinsky:1983zz}).
It will be our next task to find out differences.
Since there are two more independent parameters compared with the
Higgs inflation
and three-field system is approximated by a single-filed system,
such differences should exist  in particular 
in  the primordial non-Gaussianity in the cosmological density perturbations
which manifests itself in the CMB anisotropy \cite{Bartolo:2004if}.
We will come to address these problems elsewhere.
 
 The chiral symmetry breaking in the hidden sector produces 
 (quasi) NG bosons in the same way as in QCD. 
Due to  the Yukawa coupling of $S$
with the hidden  fermions, which explicitly breaks the chiral symmetry,
the  (quasi)  NG bosons are massive.  In contrast to  the case of QCD 
they are stable because of the unbroken vector-like flavor symmetry and 
 therefore can be DM candidates.
We have however 
realized that the full $SU(3)_V$  flavor group has to be broken
to obtain a realistic DM relic abundance.
The reason  is that
the (quasi) NG boson mass decreases as the Yukawa coupling $y$ decreases,
while the energy scale of the hidden sector $\Lambda_H$ increases
(because the scales of the visible sectors are fixed).
The only viable scenario for a realistic DM  in our model is  
the decay from inflaton $S$. 
For that to work  the Yukawa coupling should be very small to sufficiently 
suppress the decay width ($\propto y^2$), 
which
would imply that  $\Lambda_H$  would be  several orders of magnitude 
larger than $M_\text{Pl}$.
If on the other hand $SU(3)_V$ is explicitly  broken to $SU(2)_V\times U(1)$, 
we have
two Yukawa couplings $y_1=y_2$ and $y_3$.
For large $y_3\simeq 10^{-3}$ and small $y_1\simeq
10^{-14\,\sim\,-13}$, which  gives $\Lambda_H /M_\text{Pl} \simeq 10^{-1} $ and
$m_\text{DM}\,(\propto \sqrt{y_1/y_3})\simeq 10^{9\,\sim\,12}$ GeV,
the DM relic abundance can become comparable 
with the observed value.
Unfortunately, it will be impossible to directly 
detect our DM, because it is too heavy,
and the interaction with the visible sector is suppressed by $y_1$ and 
hence negligibly small.

\section*{Acknowledgments}

 J.~K.~would like to thank
J.~Kuntz, M.~Lindner, J.~Rezacek,
P.~Saake and A.~Trautner for useful and interesting
discussions.
The work of M.~A. is supported in part by the Japan Society for the
Promotion of Sciences Grant-in-Aid for Scientific Research (Grant
No. 17K05412 and No. 20H00160).
J.~K.~is partially supported by the Grant-in-Aid for Scientific Research (C) from the Japan Society for Promotion of Science (Grant No.19K03844).
J. Y. is supported by the China Scholarship Council and the Japanese Government
 (Monbukagakusho-MEXT) scholarship.

\bibliographystyle{JHEP}
\bibliography{uosref1}
\end{document}